\documentclass[floats,floatfix,showpacs,amssymb,prd,twocolumn,superscriptaddress,nofootinbib]{revtex4-1}
\usepackage{graphicx, epsfig,amssymb} 
\usepackage{epstopdf}
\usepackage{amsmath, amsfonts}
\usepackage{bm} 
\usepackage[breaklinks]{hyperref}
\usepackage{color}

\newcommand{\nn}{\nonumber}

\def\be{\begin{equation}}
\def\ee{\end{equation}}
\def\beq{\begin{eqnarray}}
\def\eeq{\end{eqnarray}}

\begin{document}

\title{\large Massive spin-2 fields on black hole spacetimes:
\\
Instability of the Schwarzschild and Kerr solutions and bounds on the graviton mass
}


\author{Richard Brito} \email{richard.brito@ist.utl.pt}
\affiliation{CENTRA, Departamento de F\'{\i}sica, Instituto Superior
  T\'ecnico, Universidade T\'ecnica de Lisboa - UTL, Avenida~Rovisco Pais
  1, 1049 Lisboa, Portugal}

\author{Vitor Cardoso}
\email{vitor.cardoso@ist.utl.pt}
\affiliation{CENTRA, Departamento de F\'{\i}sica, Instituto Superior
  T\'ecnico, Universidade T\'ecnica de Lisboa - UTL, Avenida~Rovisco Pais
  1, 1049 Lisboa, Portugal}
\affiliation{Perimeter Institute for Theoretical Physics
Waterloo, Ontario N2J 2W9, Canada}
\affiliation{Department of Physics and Astronomy, The University of Mississippi, University, MS 38677, USA.
}

\author{Paolo Pani}
\email{paolo.pani@ist.utl.pt}
\affiliation{CENTRA, Departamento de F\'{\i}sica, Instituto Superior
  T\'ecnico, Universidade T\'ecnica de Lisboa - UTL, Avenida~Rovisco Pais
  1, 1049 Lisboa, Portugal}
\affiliation{Institute for Theory $\&$ Computation, Harvard-Smithsonian
CfA, 60 Garden Street, Cambridge, MA 02138, USA}
 
\begin{abstract}
Massive bosonic fields of arbitrary spin are predicted by general extensions of the Standard Model. 
It has been recently shown that there exists a family of bimetric theories of gravity --~including massive gravity~--  which are free of Boulware-Deser ghosts at the nonlinear level. This opens up the possibility to describe consistently the dynamics of massive spin-2 particles in a gravitational field.
Within this context, we develop the study of massive spin-2 fluctuations --~including massive gravitons~-- around Schwarzschild and slowly-rotating Kerr black holes. Our work has two important outcomes. First, we show that the Schwarzschild geometry is linearly unstable for small tensor masses, against a spherically symmetric mode. Second, we provide solid evidence that the Kerr geometry is also generically unstable, both against the spherical mode and against long-lived superradiant modes. In the absence of nonlinear effects, the observation of spinning black holes bounds the graviton mass $\mu$ to be $\mu\lesssim 5\times 10^{-23} {\rm eV}$.
\end{abstract}
 
\maketitle

\section{Introduction}
The feebleness with which exotic particles--~such as those predicted in several extensions of the Standard Model~\cite{Arvanitaki:2009fg,Goodsell:2009xc,Jaeckel:2010ni} or in modified theories of gravity~\cite{Clifton:2011jh}-- couple to ordinary matter, lies at the heart of the difficulty to detect them. Extra fundamental fields may couple to Standard Model particles in various ways, which makes it challenging to exclude, or possibly detect new effects.

Fortunately, the equivalence principle guarantees that all forms of matter gravitate.
Therefore, it is no surprise that extra fundamental fields --~{\it especially} if extremely light by Standard Model standards~-- can strongly affect the dynamics of selfgravitating compact objects, such as black holes (BHs) and neutron stars. 
The equivalence principle, together with the fact that BHs are \emph{vacuum} solutions, guarantees that all forms of matter, including exotic matter, interact with BHs in the same universal way. One is thus offered the intriguing possibility of using the growing wealth of observations in high-energy astrophysics~\cite{Narayan:2005ie,Brenneman:2011wz,2013Natur.494..449R} to put physics beyond the Standard Model to the test.

There is a vast literature --which we will not attempt at summarizing-- on the gravitational interaction of fundamental scalar fields~\cite{Fujii:2003pa}. Of more direct interest to us are recent efforts to use BHs as particle-physics laboratories,
through which one can constrain the mass of the QCD axion, of stringy pseudoscalars populating the so-called axiverse~\cite{Arvanitaki:2009fg,Arvanitaki:2010sy,Pani:2012bp,Pani:2012vp}, and the hidden $U(1)$ sector of the Standard Model~\cite{Goodsell:2009xc,Jaeckel:2010ni,Pani:2012bp,Pani:2012vp}. In addition to their phenomenological relevance, such studies have revealed unexpected aspects related to the dynamics of these fields in curved spacetime.

In this paper, we take a further step in this enterprise by investigating the dynamics of massive spin-2 fields propagating on a BH spacetime.

\subsection*{Executive summary}
For the reader's convenience, we summarize here the structure of the paper and our main results.
To put our work into context, Section~\ref{sec:longintro} is devoted to a generic discussion on massive gravity~\cite{deRham:2010ik,deRham:2010kj,Hassan:2011hr} and bimetric theories~\cite{Isham:1971gm,Salam:1976as,Hassan:2011zd}, on the dynamics of spin-2 fields on curved spacetimes and their possible imprint in gravitational-wave and BH physics. We also discuss how ultralight spin-2 fields are expected to trigger strong superradiant instabilities~\cite{zeldo1,zeldo2,Press:1972zz,Cardoso:2012zn} in massive BHs.

In Section~\ref{sec:theory} we review Fierz-Pauli theory~\cite{Fierz:1939ix} for a linearized massive spin-2 field propagating on flat and curved backgrounds (see also Ref.~\cite{Hinterbichler:2011tt}). The linearized field equations for a massive spin-2 fluctuation propagating on curved spacetimes are given in Eqs.~\eqref{eqmotioncurved}--\eqref{constraint2} and we also show how they can be consistently obtained in bimetric and massive gravity.

Within this context, Sections~\ref{sec:monopoleinstability} and~\ref{sec:schwar} are devoted to a complete analysis of the linear dynamics on a Schwarzschild BH. In Sec.~\ref{sec:monopoleinstability} we focus on the monopole mode that corresponds to the scalar polarization of a massive graviton. We find a strongly unstable, spherically symmetric mode, which was also discussed very recently in Ref.~\cite{Babichev:2013}. 
Thus Schwarzschild BHs are unstable in these theories and we show that the inclusion of a cosmological constant makes the Schwarzschild-de Sitter BHs even more unstable. Furthermore, in Sec.~\ref{sec:schwar} we derive the full master equations for the axial and polar sectors. We find that the spectrum supports quasinormal modes (QNMs) and quasibound, long-lived states for any non spherically symmetric mode and we compute the spectrum numerically.

In Sec.~\ref{sec:kerr} we extend our analysis to stationary and axisymmetric BHs, namely to the Kerr metric. In general, the radial and angular part of the perturbation equations on a spinning geometry are challenging --~if possible at all~-- to separate within the standard Teukolsky approach~\cite{Berti:2009kk,PaniNRHEP2}. The same obstacle is encountered for massive spin-1 (Proca) perturbations of a Kerr BH. The problem has been recently solved within a slow-rotation framework~\cite{Kojima:1992ie,1993ApJ...414..247K,1993PThPh..90..977K} in the frequency domain~\cite{Pani:2012bp,Pani:2012vp} and also using full-fledged numerical evolutions in the time domain~\cite{Witek:2012tr}. We have extended the technique of Refs.~\cite{Pani:2012bp,Pani:2012vp} to the case of massive spin-2 perturbations [see also~\cite{Pani:2013ija} for the case of gravito-electromagnetic perturbations of Kerr-Newman BHs].

We derive the perturbations equations to first order in the BH angular momentum. In principle, this procedure can be extended to any order. To first order, the eigenvalues of the system are described by two independent sets of equations (one for each parity) and for each harmonic index. By solving the first-order equations, we have found strong evidence for the existence of unstable modes in the spectrum. This instability is different from that affecting Schwarzschild BHs and it is associated to nonspherical modes which becomes unstable above a certain BH angular momentum. The instability can be four orders of magnitude stronger than in the Proca case and up to seven orders stronger than in the massive scalar case. 
Our results provide strong indications that massive spin-2 fields trigger the strongest superradiant instability in vacuum BH solutions.

Although a second-order analysis would be necessary to describe superradiance consistently, a first-order approximation is generally sufficient to give accurate results well beyond its regime of validity~\cite{Pani:2012bp}. Including second-order effects would be an important --~and technically challenging~-- extension of our work.
The unstable, spherically-symmetric mode active for Schwarzschild BHs is unaffected by rotation, at first order. Thus, we present {\it two}
mechanisms by which Kerr BHs are rendered unstable in massive theories of gravity.

Several technicalities are discussed in the Appendices and in publicly available {\scshape Mathematica} notebooks~\cite{webpage}.
In Appendix~\ref{app:ana} we generalize Detweiler's calculation of the unstable massive scalar modes of a Kerr BH~\cite{Detweiler:1980uk} to the dipolar axial sector of massive spin-2 fields to first order in the BH angular momentum. 

We conclude in Sec.~\ref{sec:conclusion}, with some phenomenological implications and with possible future extensions of our results.

\section{Massive spin-2 fields and strong gravity}~\label{sec:longintro}
\subsection{Massive gravitons?}
Higher-spin fields are predicted to arise in several contexts~\cite{Bouatta:2004kk,Sorokin:2004ie,Sagnotti:2011qp}. The motivation to investigate their gravitational dynamics is twofold.
The first reason is conceptual and is tied to a renewed interest in massive gravity and bimetric theories of gravity. It is known since the work of Fierz and Pauli that at the linear level there is only one ghost- and tachyon-free, Lorentz-invariant mass term that describes the five polarizations of a massive spin-2 field on a flat background~\cite{Fierz:1939ix}. 
However, in the zero-mass limit the Fierz-Pauli theory does not recover linear general relativity due to the existence of extra degrees of freedom introduced by the graviton mass. In the massless limit the helicity-0 state maintains a finite coupling to the trace of the source stress-energy tensor, modifying the Newtonian potential and hence yielding predictions
which differ from the massless graviton theory~\cite{Iwasaki:1971uz,VanNieuwenhuizen:1973qf,Carrera:2001pj,Hinterbichler:2011tt}, rendering the theory inconsistent with observations. This is known as the vDVZ~discontinuity~\cite{vanDam:1970vg,Zakharov:1970cc}. 

To overcome this difficulty Vainshtein~\cite{Vainshtein:1972sx} argued that the discontinuity present in the Fierz-Pauli theory is an artifact of the linear theory, and that the full nonlinear theory has a smooth limit for $m_g\equiv\hbar\mu\to 0$. He found that around any massive object of mass $M$, there is a new length scale known as the Vainshtein radius, $r_V\sim\left(M/(m_g^4 M_p^2)\right)^{1/5}$. 
The nonlinearities begin to dominate at $r\lesssim r_V$ invalidating the predictions made by the linear theory. This is due to the fact that at high energies the helicity-$0$ mode of the graviton, responsible for the discontinuity, is strongly coupled to itself and becomes weakly coupled to external sources. However, it was believed until recently that Lorentz-invariant nonlinear massive gravity theories were doomed to fail due to the (re)appearance of a ghost-like sixth degree of freedom~\cite{Boulware:1973my}. This was studied by Boulware and Deser who showed that in nontrivial backgrounds there are 6 degrees of freedom, where the extra degree of freedom was shown to be a ghost scalar, known as the Boulware-Deser ghost. 

More recently, a two-parameter family of nonlinear generalizations of the linear Fierz-Pauli theory was proposed by de Rham, Gabadadze and Tolley~\cite{deRham:2010ik,deRham:2010kj,Hassan:2011hr} and it is usually referred to as ``nonlinear massive gravity'' [see Ref.~\cite{Hinterbichler:2011tt} for a review].  When linearized on a flat background, nonlinear massive gravity has so far proved to be ghost-free (but see Refs.~\cite{2013arXiv1302.4367C} for recent counterarguments and Ref.~\cite{Burrage:2012ja} for some tight constraints on the theory in the decoupling limit). The extension of the theory to generic nonflat backgrounds appears to be also ghost-free~\cite{Hassan:2011tf,Hassan:2011vm,Hassan:2011ea}. On the other hand, it has been recently shown that the very same combination that removes the Boulware-Deser ghost is also responsible for the existence of superluminal shock-wave solutions which render the theory acausal~\cite{Deser:2012qx}. 

Furthermore, the healthy interaction term that prevents the theory to propagate ghosts has been also generalized to bimetric theories of gravity, i.e. to theories which propagate two dynamical spin-2 fields~\cite{Isham:1971gm,Salam:1976as,Hassan:2011zd}. These theories can also describe a massive spin-2 field coupled to standard Einstein gravity~\cite{Hassan:2012wr} and they reduce to nonlinear massive gravity when one of the fields is nondynamical~\cite{Baccetti:2012bk}.  
%

\subsection{Gravitational-wave searches and astrophysics}

The second motivation to investigate massive spin-2 fields is of a more practical and phenomenological nature. Advanced gravitational-wave detectors will begin operation in a couple of years
and the first direct detection of a graviton on Earth is expected to take place within the next decade. Current constraints
on the graviton mass from pulsar observations already provide compelling evidence that gravitational waves {\it are} indeed emitted when two objects merge~\cite{Will:2005va}. A hypothetical massive graviton would affect the decay rate of the orbiting pulsar~\cite{Damour:1990wz,Goldhaber:2008xy}. The Hulse-Taylor pulsar provides a stringent limit on the mass of the graviton~\cite{Finn:2001qi}, $\mu\lesssim 7.6\times 10^{-20} {\rm eV}$
\footnote{Note however that the theory considered in Ref~\cite{Finn:2001qi} does not satisfy the Fierz-Pauli tuning and hence it contains a ghost. It would be interesting to repeat such calculation for viable theories. In this case however, the Vainshtein mechanism discussed in the main text may prevent a consistent linear analysis.}.

However, even with these tight constraints in place, the Yukawa-like potential of a hypothetical graviton mass would be responsible for a 
deformation of the gravitational-wave signal during its journey from the source to the observer. In other words, a small graviton mass {\it may} not affect the inspiral of a binary system to a significant extent (including the changes in period of binary pulsar),
but introduces nontrivial dispersion which acts over several Compton wavelengths, $\sim \mu^{-1}$. This peculiar effect can leave a signature in the gravitational waveform. Because any putative gravitational-wave detections will occur with very low signal-to-noise-ratio, an accurate knowledge
of these effects may be important, in the sense that accurate templates are required to detect extra polarizations without introducing bias~\cite{Will:1997bb,Chatziioannou:2012rf} [see Ref.~\cite{Yunes:2013review} for a recent review]. 

In summary, gravitational waveforms for inspiralling objects emitting massive gravitons are necessary. There are several ways to deal with this
problem, e.g., full nonlinear simulation, slow-motion expansions or perturbative expansions around some background.
We will initiate here the latter, by understanding how small vacuum fluctuations behave in bimetric theories and massive gravity.
As a by-product, we are able to understand stability properties of BHs in these theories and begin to understand
how gravitational waveforms differ from general relativity [see also Ref.~\cite{DeFelice:2013nba} for a recent attempt].

\subsection{Massive gravitons and the Gregory-Laflamme instability}
In a very recent paper~\cite{Babichev:2013}\footnote{Ref.~\cite{Babichev:2013} appeared while our work was on its last stages.}, Babichev and Fabbri showed
that the mass term for the graviton can be interpreted as a Kaluza-Klein
momentum of a four-dimensional Schwarzschild BH extended into a flat higher dimensional spacetime.
Such ``black string'' spacetimes are known to be unstable against long-wavelength perturbations,
or in other words, against low-mass perturbations, which are spherically symmetric on the four-dimensional subspace. 
This is known as the Gregory-Laflamme instability \cite{Gregory:1993vy,Kudoh:2006bp}, which in turn is the analog of a Rayleigh-Plateau instability of fluids~\cite{Cardoso:2006ks,Camps:2010br}. Based on these results, Ref.~\cite{Babichev:2013} pointed out that massive tensor perturbations on a Schwarzschild BH in massive gravity and bimetric theories would generically give rise to
a (spherically symmetric) instability. In the following we confirm these results within a more generic framework and extend them to generic modes and to the case of Schwarzschild-de Sitter BHs.

One of the important open questions is the end-state of such instability. For black strings, there is reasonable evidence that
break-up occurs \cite{Lehner:2010pn}. But the spacetimes we deal with are spherically symmetric, and so is the unstable mode. 
A possible end-state is a spherically symmetric BH endowed with a graviton cloud (see e.g. Ref.~\cite{Volkov:2012wp}). An analysis of the nonlinear equations in case of spherical symmetry is left for future work.

\subsection{Massive bosons and BH superradiance}
The interaction of generic bosonic fields with spinning BHs gives rise to interesting phenomena, related to 
{\rm BH superradiance}~\cite{zeldo1,zeldo2,Press:1972zz,Cardoso:2012zn}. Due to the dissipative nature of the BH horizon and to the existence of negative-energy states in the ergoregion of a spinning BH, low-frequency $\omega$ monochromatic bosonic waves scattered off 
rotating BHs are amplified whenever the following condition is met,
\be
\omega<m \Omega_H\,,
\ee
where $\Omega_H$ is the angular velocity of the BH horizon and $m$ is an integer characterizing the azimuthal dependence of the wave. The extra energy deposited in the wavepacket's amplitude is extracted
from the BH, which spins down. 

Superradiance is prone to very interesting ``side-effects,'' such as BH bombs \cite{Press:1972zz,Cardoso:2004nk}, floating orbits
\cite{Cardoso:2011xi,Cardoso:2012zn,Brito:2012gw} and BH instabilities \cite{Detweiler:1980uk,Cardoso:2004hs,Cardoso:2005vk,Cardoso:2007az,Dolan:2007mj,Yoshino:2012kn,Pani:2012vp,Witek:2012tr,Dolan:2012yt,Hod:2012zza}
(for a review see Ref.~\cite{Cardoso:2013zfa}).

The amount of energy extracted through superradiance strongly depends on the spin of the field. Massless spin-2 (gravitational) waves can be amplified $\sim300$ times more than scalar waves. Superradiance scattering for massive waves with nonvanishing spin is much more involved, due to spin-spin coupling effects~\cite{Pani:2012bp}. However, a generic expectation is that superradiant instabilities triggered by massive bosons are more effective for higher spin. Finally, even in the scalar case superradiant effects might be enormously amplified due to the interaction with ordinary matter~\cite{Cardoso:BHST}.

We are particularly interested in superradiance-triggered BH instabilities which are sustained by massive fields.
Ultralight bosons have received widespread attention recently as they are found in several extensions of the Standard Model,
for instance in the string axiverse scenario~\cite{Arvanitaki:2009fg,Arvanitaki:2010sy} where a plethora of massive pseudo-scalar fields called axions covers each decade of mass range down to the Hubble scale and fields with $10^{-22} {\rm eV}<m_s<10^{-10}{\rm eV}$ are of particular interest for BH physics~\cite{Kodama:2011zc}. In parallel, massive hidden $U(1)$ vector fields also arise in extensions of the standard model~\cite{Goodsell:2009xc,Jaeckel:2010ni,Camara:2011jg,Goldhaber:2008xy}, highlighting the importance of understanding the physics of such fields around BHs.

Superradiant instabilities were studied extensively for scalar fields both in the frequency- and in the 
time-domain~\cite{Detweiler:1980uk,Cardoso:2004nk,Cardoso:2005vk,Dolan:2007mj,Witek:2012tr,Dolan:2012yt,Cardoso:BHST}.
The non-separability of the field equations for a massive vector field in a Kerr background has hampered its study for decades (see for instance
Ref.~\cite{Rosa:2011my} for some references on the nonrotating case). Very recently however, progress has been made. 
In the frequency domain slow-rotating expansions were used to prove that massive vectors {\it are}
superradiantly unstable~\cite{Pani:2012bp,Pani:2012vp}, these results were confirmed using evolutions of wavepackets around Kerr BHs ~\cite{Witek:2012tr}. It was shown that the massive vector field instability can be orders of magnitude stronger than the massive scalar field.

The instability is regulated by two parameters, the BH spin $a/M$ and the dimensionless parameter $M\mu$ (in units $G=c=1$), where $M$ is the
BH mass and $m_g=\mu\hbar$ is the bosonic field mass. For ultralight scalar fields around massive BHs, the instability timescale can be of the order of seconds for solar-mass BHs and of the order of hundreds years for a supermassive BH with $M\sim 10^9 M_\odot$~\cite{Detweiler:1980uk,Arvanitaki:2009fg,Arvanitaki:2010sy}, typically much shorter than the evolution timescale of astrophysical objects. The instability timescale for spin-1 massive fields can be up to three orders of magnitude shorter~\cite{Pani:2012bp,Pani:2012vp,Witek:2012tr}. 
To summarize, this mechanism can be very efficient for extraction of angular momentum away from the BH.
As a consequence, observations of massive spinning BHs can effectively be used to impose bounds on ultralight boson masses~\cite{Pani:2012bp}.


\subsection{Framework}

We wish to describe two different cases: i) the interaction of a generic massive spin-2 field with standard gravity, that is, we consider the massive tensor as a probe field propagating on a geometry which solves Einstein equations; ii) the linearized dynamics of a massive graviton as it emerges in nonlinear massive gravity. It turns out that both cases can be described consistently within a common framework.

More specifically, we consider the action for two tensor fields, $g_{\mu\nu}$ and $f_{\mu\nu}$, with a ghost-free nonlinear interaction between them (cf. Eq.~\eqref{biaction} below). This class of theories is usually referred to as ``bimetric gravity''~\cite{Isham:1971gm,Salam:1976as,Hassan:2011zd}. 
The fluctuations of the two dynamical metrics can be separated and describe two interacting gravitons, one massive and one massless.

Nonlinear massive gravity~\cite{deRham:2010kj,deRham:2010ik,Hassan:2011zd} is obtained from the bimetric theory in the limit where the field $f_{\mu\nu}$ becomes nondynamical, i.e. taking $M_f\to0$ in Eq.~\eqref{biaction} and considering $f_{\mu\nu}$ as a given auxiliary field~\cite{Baccetti:2012bk}. 
In this limit,  $f_{\mu\nu}$ can be interpreted as a background metric in which the linearized massive fluctuations $h^{(m)}_{\mu\nu}$ propagate. On the other hand, $g_{\mu\nu}$ is a solution of the full non-linear field equations such that we have $g_{\mu\nu}=f_{\mu\nu}+h^{(m)}_{\mu\nu}$.

A crucial point is to identify the background solution over which the massive tensor perturbations propagate. Linearization of massive gravity is typically considered around a flat, Minkowski background. Here instead we wish to describe the linearized dynamics around a nonlinear vacuum solution, i.e. a BH geometry. Regular, nonlinear, solutions in bimetric and massive gravity are challenging to find and they might exhibit a rich structure~\cite{Deffayet:2011rh,Berezhiani:2011mt,Banados:2011hk,Cai:2012db}. 
In bimetric theories new curvature invariants, such as $I=g^{\mu\nu}f_{\mu\nu}$, can become singular at the horizon. It was shown that the only way to avoid a singular horizon is to require both metrics to have coincident horizons~\cite{Deffayet:2011rh,Banados:2011hk}. 
The same arguments were used to show that regular BHs can exist in massive gravity theories with a flat nondynamical metric provided
at least one of the metrics is non-diagonal (or non-stationary and axisymmetric) when written in the same coordinate patch~\cite{Deffayet:2011rh}. 

In massive gravity the diffeomorphism of general relativity is broken, so in principle one is not allowed to change coordinates to avoid this problem. This implies that, assuming a flat background, BH solutions in Schwarzschild coordinates must have a component $g_{{\rm tr}}$ to avoid a singular horizon. This component implies a time-dependence and nonzero energy flux $T_{{\rm tr}}$ near the horizon, which might even lead to the disappearance of BHs in this theory~\cite{Mirbabayi:2013sva}. Due to the Yukawa-like potential the BH gravitational field is screened by a negative energy density which is accreted by the BH because of the ingoing flux $T_{{\rm tr}}$ leading to a decrease of the BH mass. Although the timescale should be much longer than the Hubble time (and hence astrophysically irrelevant), it seems to be an anomaly of massive gravity. 

To avoid dealing with such problems, we consider the special case in which the background solutions are the same as in general relativity. In bimetric theories this can be accomplished by taking the two metrics to be proportional, $f_{\mu\nu}=C^2g_{\mu\nu}$, as discussed in detail below [see also Ref.~\cite{Hassan:2012wr}]. This choice also avoids the singular horizon problem, as the two metrics have the same horizon. The linearized equations describing the fluctuations of the two metrics can be easily decoupled and they describe one massless graviton (which is described by usual linearized Einstein dynamics), and a massive graviton which is described by the Fierz-Pauli theory on a curved background~\cite{Hassan:2012wr,Volkov:2013roa}.

On the other hand, in the limit of massive gravity this is equivalent of taking the nondynamical metric as being the BH spacetime instead of the usual flat spacetime. Although perfectly consistent with the field equations, this choice seems somewhat unnatural and other nonlinear background metrics can be considered [cf. Ref.~\cite{Volkov:2013roa} for a recent review].
The fluctuations of the physical metric $g_{\mu\nu}$ propagate on a nonlinear BH background $f_{\mu\nu}$ and they are also described by Fierz-Pauli theory. 



\section{Linearized Massive Gravity on curved spacetime}\label{sec:theory}
\subsection{The Fierz-Pauli tuning in flat spacetime}
Let us start by reviewing the classical Fierz-Pauli theory describing a massive spin-2 field in four-dimensional flat spacetime. The action is given by~\cite{Fierz:1939ix}
\begin{align}\label{FPaction}
S_{FP}&=\frac{1}{16\pi G}\int\,d^4x\,\left[-\frac{1}{2}\partial_{\lambda}h_{\mu\nu}\partial^{\lambda}h^{\mu\nu}+\partial_{\mu}h_{\nu\lambda}\partial^{\nu}h^{\mu\lambda}\right.\nn\\
&\left.-\partial_{\mu}h^{\mu\nu}\partial_{\nu}h+\frac{1}{2}\partial_{\lambda}h\partial^{\lambda}h-\frac{\mu^2}{2}\left(h_{\mu\nu}h^{\mu\nu}-\kappa h^2\right)\right]\,, \nn
\end{align}
where $h=\eta^{\mu\nu}h_{\mu\nu}$ is the trace of the symmetric tensor field $h_{\mu\nu}$, $\eta^{\mu\nu}$ is the Minkowski metric, $\kappa$ is an arbitrary constant, and $\mu$ is the graviton mass. When $\mu=0$, the action reduces to the linearized Einstein-Hilbert action. When $\mu\neq0$, the mass term violates the diffeomorphism invariance of general relativity, i.e., this action is not invariant under infinitesimal transformations of the form
\be
\delta h_{\mu\nu}=\partial_{\mu}\xi_{\nu}(x)+\partial_{\nu}\xi_{\mu}(x)\,.
\ee

The equations of motion are given by [see Ref.~\cite{Hinterbichler:2011tt} for a review]
\begin{align}\label{eqmotion}
\frac{\delta S}{\delta h_{\mu\nu}}&=\Box h_{\mu\nu}-\partial_{\lambda}\partial_{\mu}h^{\lambda}_{\nu}-\partial_{\lambda}\partial_{\nu}h^{\lambda}_{\mu}+
\eta_{\mu\nu}\partial_{\lambda}\partial_{\sigma}h^{\lambda\sigma} \nn\\
&+\partial_{\mu}\partial_{\nu}h-\eta_{\mu\nu}\Box h-\mu^2\left(h_{\mu\nu}-\kappa \eta_{\mu\nu}h\right)=0\,. 
\end{align}
Acting with $\partial^{\mu}$ on \eqref{eqmotion} we find the constraint
\be
\label{harmonic}
\partial^{\nu}h_{\nu\mu}-\kappa\partial_{\mu}h=0\,.
\ee
Note that for $\kappa={1}/{2}$ this corresponds to the harmonic gauge in linearized general relativity. Plugging this back into the field equations and taking the trace, we find
\be
\label{trace}
2(1-\kappa)\Box h+(1-4\kappa)\mu^2 h=0\,.
\ee
Substituting the trace condition, Eq.~\eqref{eqmotion} reads
\be
\label{eqmotion2}
(\Box-\mu^2)h_{\mu\nu}=(2\kappa-1)\left[\partial_{\mu}\partial_{\nu}h+\frac{1}{2}\eta_{\mu\nu}\mu^2 h\right]\,.
\ee
For massive spin-2 particles we must have $2s+1=5$ degrees of freedom. The only choice for the constant $\kappa$ that describes a single massive graviton is the Fierz-Pauli tuning, $\kappa=1$~\cite{Fierz:1939ix}.
In this case, the full set of linearized equations reads:
\be
\label{eqmotion3}
(\Box-\mu^2)h_{\mu\nu}=0\,,\qquad \partial^{\mu}h_{\mu\nu}=0\,,\qquad h=0\,.
\ee
On the other hand, for $\kappa\neq 1$ the theory propagates 6 degrees of freedom. The extra polarization comes from a scalar ghost (a scalar with negative kinetic energy) of mass $m_{{\rm ghost}}^2=-\frac{1-4\kappa}{2(1-\kappa)}\mu^2$, which arises from the trace equation~\eqref{trace}. The ghost mass approaches infinity as the Fierz-Pauli tuning is approached, so that the ghost decouples in this limit. 


\subsection{Massive spin-2 particles on curved spacetimes}
Let us now generalize the equations of motion for massive spin-2 particles on a curved background~\cite{Bengtsson:1994vn,Buchbinder:1999ar,Hassan:2012wr}. The more general ghost-free action of two interacting spin-2 fields, without matter couplings, is given by~\cite{Hassan:2011zd}
\be
\-\-S=\int d^4x\sqrt{|g|}\left[M_g^2R_g+M_f^2\sqrt{\frac{f}{g}}R_f-2M_v^4V\left(g,f\right)\right]\,, \label{biaction}
\ee
where $R_g$ and $R_f$ are the Ricci scalars corresponding to $g_{\mu\nu}$ and $f_{\mu\nu}$, respectively; $M_g^{-2}={16\pi G}$, $M_f^{-2}={16\pi \mathcal{G}}$ are the corresponding gravitational couplings, and $M_v$ is written in terms of $M_g$, $M_f$ and of the parameters of the potential term. The quantities $f,g$ denote the determinant of the respective metric. 
There is a unique prescription for the latter in terms of only five interaction terms which is free from the Boulware-Deser ghosts on generic backgrounds. We schematically denote the potential as
\be
\label{potential}
V\equiv\sum_{n=0}^4\,\beta_n V_n\left(\gamma\right)\,, \quad \gamma^{\mu}\,_{\nu}=\left(\sqrt{g^{-1}f}\right)^{\mu}\,_{\nu}
\ee
where $\beta_i$ are coupling constants. The precise form of the potentials $V_n$ is not crucial here and we refer to the original papers~\cite{deRham:2010kj,deRham:2010ik,Hassan:2011zd}.


Although the action~\eqref{biaction} describes a vacuum bimetric theory, it reduces to massive gravity in the limit $M_f\to 0$, in which case the kinetic term of the metric $f_{\mu\nu}$ vanishes and the field is taken to be auxiliary~\cite{Baccetti:2012bk}. 

From the action~\eqref{biaction} we find two sets of Einstein's equations for $g_{\mu\nu}$ and $f_{\mu\nu}$ 
\beq\
R_{\mu\nu}(g)-\frac{1}{2}g_{\mu\nu}R(g)+\frac{M_v^4}{M_g^2}\mathcal{T}^g_{\mu\nu}(\gamma)&=&0\,, \label{field_eqs1} \\
R_{\mu\nu}(f)-\frac{1}{2}f_{\mu\nu}R(f)+\frac{M_v^4}{M_f^2}\mathcal{T}^f_{\mu\nu}(\gamma)&=&0\,, \label{field_eqs2}
\eeq
where the ``graviton'' stress-energy tensors $\mathcal{T}_{\mu\nu}^g$ and $\mathcal{T}_{\mu\nu}^f$ depend on $\gamma^\mu\,_\nu$ and are defined, e.g., in Ref.~\cite{Hassan:2012wr}.
%

Since we want to consider a BH geometry as background, we first need to 
find a BH solution of the field equations. As previously discussed, this is a challenging and controversial issue [see Ref.~\cite{Volkov:2013roa} for a recent survey of hairy BHs in massive gravity].

Here we make the simplest choice and consider two proportional \emph{background} metrics $\bar{f}_{\mu\nu}=C^2\bar{g}_{\mu\nu}$ (we use the bar notation to denote background quantities). Remarkably, in this case the solutions coincide with those of general relativity. Indeed, Eqs.~\eqref{field_eqs1} and \eqref{field_eqs2} reduce to~\cite{Hassan:2012wr} 
\beq
\label{eqs_pro}
\bar{R}_{\mu\nu}-\frac{1}{2}\bar{g}_{\mu\nu}\bar{R}+\Lambda_g \bar{g}_{\mu\nu}&=&0\,,\nn\\
\bar{R}_{\mu\nu}-\frac{1}{2}\bar{g}_{\mu\nu}\bar{R}+\Lambda_f \bar{g}_{\mu\nu}&=&0\,,
\eeq
which are just two copies of Einstein's equations with two different cosmological constants. The latter are written in terms of the parameters of the interaction potentials and of the gravitational couplings~\cite{Hassan:2012wr}. Furthermore, consistency of the background equations requires $\Lambda_g=\Lambda_f$, which translates into a quartic algebraic equation for the constant $C$. 
Classical no-hair theorems of general relativity guarantee that the most general stationary BH solution in vacuum and with a cosmological constant is the Kerr-(Anti) de Sitter metric. Therefore, when $\Lambda_g=\Lambda_f>0$ the fields $g_{\mu\nu}$ and $f_{\mu\nu}$ describe two identical Kerr-de Sitter BHs.

Since we are interested in local physics near massive BHs, we shall consider $\Lambda_g\approx0\approx\Lambda_f$. This condition can be satisfied exactly by requiring a fine tuning of the interaction couplings~\cite{Hassan:2012wr}. Alternatively, even without fine tuning, realistic values of the cosmological constant should not play any role in describing local physics at the scale of astrophysical compact objects. Therefore, we can safely neglect those terms and focus on asymptotically-flat Kerr BHs as background solutions. In Boyer-Lindquist coordinates, these are described by the line element:
\begin{align}
\label{Kerr}
&ds_{\rm Kerr}^2=-\left(1-\frac{2Mr}{\Sigma}\right)dt^2+\frac{\Sigma}{\Delta}dr^2-\frac{4M^2r}{\Sigma}\tilde{a}\sin^2\theta d\phi dt \nn\\
&+\Sigma d\theta^2+\left[(r^2+M^2\tilde{a})\sin^2\theta+\frac{2M^3 r}{\Sigma}\tilde{a}^2\sin^4\theta \right]d\phi\,,
\end{align}
where $\Sigma=r^2+M^2\tilde{a}^2\cos^2\theta$, $\Delta=(r-r_+)(r-r_-)$, $r_\pm=M(1\pm \sqrt{1-\tilde{a}})$ and $\tilde{a}=J/M^2$.
This spacetime describes a rotating BH with mass $M$ and angular momentum $J$ in $G=c=1$ units.

Let us now consider fluctuations around the background metrics:
\begin{align}\label{pert}
g_{\mu\nu}&=\bar{g}_{\mu\nu}+\frac{1}{M_g}\delta g_{\mu\nu}\,,\\
f_{\mu\nu}&=C^2\bar{g}_{\mu\nu}+\frac{C}{M_f}\delta f_{\mu\nu}\,.
\end{align}
Note that the perturbations are generically independent, $\delta g_{\mu\nu}\neq\delta f_{\mu\nu}$.
From Eqs.~\eqref{field_eqs1}-\eqref{field_eqs2}, the linearized field equations read
\begin{align}
\label{eq_g}
&\bar{\mathcal{E}}^{\rho\sigma}_{\mu\nu}\delta g_{\rho\sigma}-\frac{M_v^4B}{M_g}\bar{g}_{\mu\rho}\left(\delta S^{\rho}\,_{\nu}-\delta_{\nu}^{\rho}\delta S^{\sigma}\,_{\sigma}\right)=0\,,\\
\label{eq_f}
&\bar{\mathcal{E}}^{\rho\sigma}_{\mu\nu}\delta f_{\rho\sigma}+\frac{M_v^4B}{C M_f}\bar{g}_{\mu\rho}\left(\delta S^{\rho}\,_{\nu}-\delta_{\nu}^{\rho}\delta S^{\sigma}\,_{\sigma}\right)=0\,,
\end{align}
where $B$ is a constant~\cite{Hassan:2012wr},
\be
\delta S^{\rho}\,_{\nu}=\frac{\bar{g}^{\rho\mu}}{2M_f}\left(\delta f_{\mu\nu}-C\frac{M_f}{M_g}\delta g_{\mu\nu}\right)\,,
\ee
and $\bar{\mathcal{E}}^{\rho\sigma}_{\mu\nu}$ is the operator representing the linearized Einstein equations in curved spacetimes:
\begin{align}
&\bar{\mathcal{E}}^{\rho\sigma}_{\mu\nu}=-\frac{1}{2}\left[\delta^{\rho}_{\mu}\delta^{\sigma}_{\nu}\bar{\Box}+\bar{g}^{\rho\sigma}\bar{\nabla}_{\mu}\bar{\nabla}_{\nu}
-\delta^{\rho}_{\mu}\bar{\nabla}^{\sigma}\bar{\nabla}_{\nu}\right.\nn\\
&\left.-\delta^{\rho}_{\nu}\bar{\nabla}^{\sigma}\bar{\nabla}_{\mu}-\bar{g}_{\mu\nu}\bar{g}^{\sigma\rho}\bar{\Box}+\bar{g}_{\mu\nu}\bar{\nabla}^{\rho}\bar{\nabla}^{\sigma}\right]\,,\label{operator}
\end{align}
where we already assumed $\Lambda_g=0=\Lambda_f$.

Taking appropriate linear combinations of the metric fluctuations,
\begin{align}
h^{(0)}_{\mu\nu}&=\frac{M_g\delta g_{\mu\nu}+C\,M_f\delta f_{\mu\nu}}{\sqrt{C^2M^2_f+M^2_g}}\,,\\
h^{(m)}_{\mu\nu}&=\frac{M_g\delta f_{\mu\nu}-C\,M_f\delta g_{\mu\nu}}{\sqrt{C^2M^2_f+M^2_g}}\,,
\end{align}
the linear equations decouple:
\begin{align}
\label{eq_0}
&\bar{\mathcal{E}}^{\rho\sigma}_{\mu\nu}h^{(0)}_{\rho\sigma}=0\,,\\
\label{eq_m}
&\bar{\mathcal{E}}^{\rho\sigma}_{\mu\nu}h^{(m)}_{\rho\sigma}+\frac{\mu^2}{2}\left(h^{(m)}_{\mu\nu}-\bar{g}_{\mu\nu}h^{(m)}\right)=0\,.
\end{align}
From the equations above, it is clear that the theory describes two spin-2 fields, $h^{(0)}_{\mu\nu}$ and $h^{(m)}_{\mu\nu}$. The former is massless and it is described by the linearized Einstein-Hilbert action, whereas the latter has a Fierz-Pauli mass term defined as
\be\label{mass_g}
\mu^2=M_v^4(C\beta_1+2C^2\beta_2+C^3\beta_3)\left(\frac{1}{C^2M_f^2}+\frac{1}{M_g^2}\right)\,.
\ee
Note that not all parameter $\beta_i$ in the equations above are independent~\cite{Hassan:2011tf}.

What we have discussed so far is valid for bimetric theories~\eqref{biaction}. It is worth stressing that linearized massive gravity can be recovered taking the limit $\delta f_{\mu\nu}\to 0$ and $M_f\to 0$ in Eq.~\eqref{pert} such that $\delta f_{\mu\nu}/M_f\to 0$. In this limit only Eq.~\eqref{eq_g} survives as a dynamical equation. In the massive gravity limit, this equation can be written in the same form as in Eq.~\eqref{eq_m} for the perturbation $\delta g_{\mu\nu}$, but with a mass term 
\be
\mu=\sqrt{BC}M_v^2/M_g\,.
\ee
Therefore, also in this case the theory describes a massive graviton propagating in the curved background $\bar{g}_{\mu\nu}\equiv \bar{f}_{\mu\nu}/C^2$.

We have just proved that in both cases (bimetric theories and massive gravity) the linearized equations describing a massive spin-2 field on a curved spacetime are described by an equation of the form~\eqref{eq_m}. In the case of bimetric theory one also has Eq.~\eqref{eq_0}, which we ignore since it describes a standard massless graviton and it is decoupled.

In flat spacetime, the equations of motion~\eqref{eq_m} reduce to Eq.~\eqref{eqmotion} whereas, on curved background they reduce to the system:
\beq
\label{eqmotioncurved}
&&\bar\Box h_{\mu\nu}+2 \bar R_{\alpha\mu\beta\nu} h^{\alpha\beta}-\mu^2 h_{\mu\nu}=0\,,\\
\label{constraint1}
&&\bar\nabla^{\mu}h_{\mu\nu}=0\,,\\
\label{constraint2}
&&h_{\mu}\,^{\mu}=0\,,
\eeq
where, here and in the following, we have suppressed the superscript ``$(m)$'' for simplicity.
This set of equations can be shown to be the only one that consistently describes a massive spin-2 coupled to gravity in generic backgrounds~\cite{Buchbinder:1999ar}. In the rest of this paper we will investigate Eqs.~\eqref{eqmotioncurved}--\eqref{constraint2} on a BH background.

%
%
%
\begin{figure*}[htb]
\begin{center}
\begin{tabular}{cc}
\epsfig{file=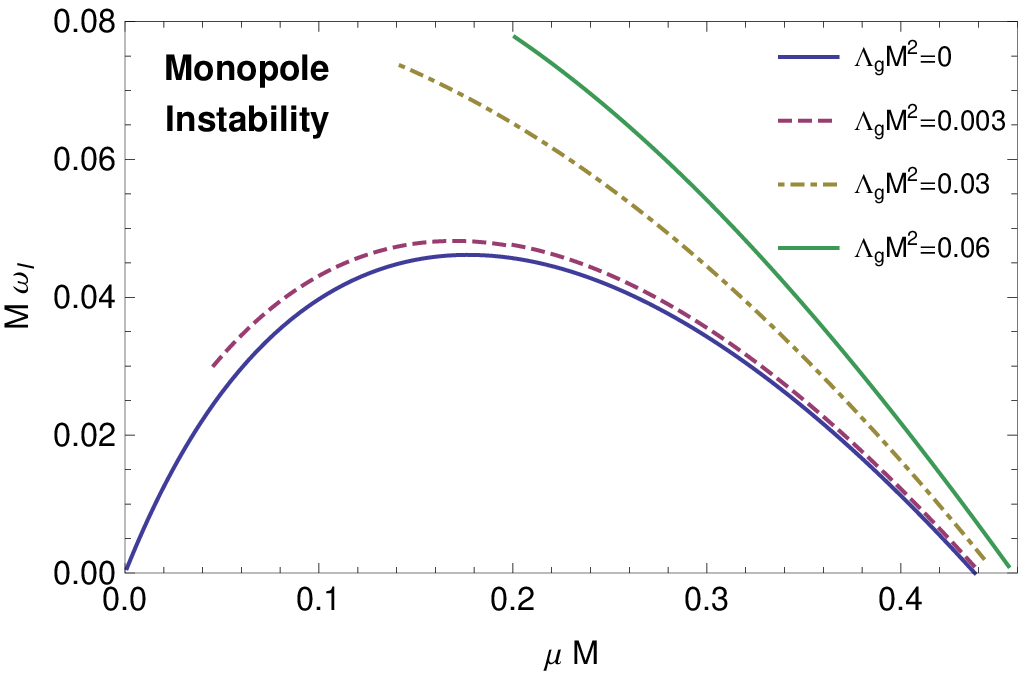,width=7.5cm,angle=0,clip=true}
\epsfig{file=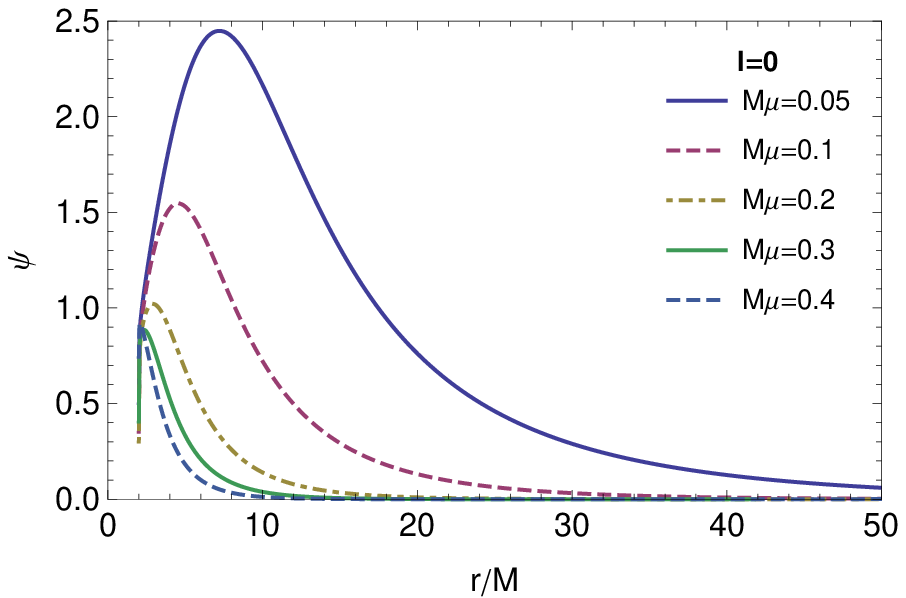,width=7.5cm,angle=0,clip=true}\\
\end{tabular}
\caption{Details of the instability of Schwarzschild (de Sitter) BHs against spherically symmetric polar modes of a massive spin-2 field. The left panel shows the inverse of the instability timescale
$\omega_I=1/\tau$ as a function of the graviton mass $\mu$ for different values of the cosmological constant $\Lambda_g=\Lambda_f$, including the asymptotically flat case $\Lambda_g=0$. Curves are truncated when the Higuchi bound is reached $\mu^2=2\Lambda_g/3$~\cite{Higuchi1987397}. For any value of $\Lambda_g$, unstable modes exist in the range $0<M\mu\lesssim 0.47$, the upper bound being only mildly sensitive to $\Lambda_g$. The right panel shows some eigenfunctions in the asymptotically flat case. The eigenfunctions decay exponentially at spatial infinity and are
progressively peaked closer and closer to the BH horizon for masses close to the threshold mass $M\mu\sim 0.43$.
\label{fig:GL}}
\end{center}
\end{figure*}
%
\section{Instability of black holes against spherically symmetric fluctuations}\label{sec:monopoleinstability}

We start by showing that Schwarzschild BHs are generically unstable against spherically symmetric perturbations~\cite{Babichev:2013}.
This is a generic and strong instability, as we will show. 
To lay the necessary framework, consider a generic tensor field $h_{\mu\nu}$ in a Schwarzschild background. Due to spherical symmetry, the tensor field $h_{\mu\nu}$ can be conveniently decomposed in a complete basis of tensor spherical harmonics~\cite{Regge:1957td,1970JMP....11.2203Z}. Furthermore, the perturbation variables are classified as ``polar'' or ``axial'' depending on how they transform under parity inversion ($\theta\to \pi-\theta$, $\phi\to \phi+\pi$). Polar perturbations are multiplied by $(-1)^l$ whereas axial perturbations pick up the opposite sign $(-1)^{l+1}$. We refer the reader to Refs.~\cite{Berti:2009kk,Zerilli:1971wd} for further terminology used in the literature. 

We decompose the spin-2 perturbation in Fourier space as follows:
%
\begin{align}
\label{decom}
h_{\mu\nu}(t,r,\theta,\phi)&=\sum_{l,m}\int_{-\infty}^{+\infty}e^{-i\omega t}\left[h^{{\rm axial},lm}_{\mu\nu}(\omega,r,\theta,\phi)\right.\nn\\
&\left.+h^{{\rm polar},lm}_{\mu\nu}(\omega,r,\theta,\phi)\right]d\omega\,.
\end{align}
where $h^{{\rm axial},lm}_{\mu\nu}$ and $h^{{\rm polar},lm}_{\mu\nu}$ are explicitly given in Appendix~\ref{app:decomposition}. In a spherically symmetric background, the field equations do not depend on the azimuthal number $m$ and they are also decoupled for each harmonic index $l$. In addition, perturbations with different harmonic opposite parity decouple from each other.

The details of the perturbation equations are provided in Appendix  \ref{app:decomposition}.
In this section, we are only interested in the $l=0$ polar sector. The perturbations $G$, $\eta_0$ and $\eta_1$ as given in Eq.~\eqref{evenpart} are not defined for $l=0$ because their angular dependence is vanishing. The remaining dynamical variables can be recast into a simple monopole equation. First, we use the constraints \eqref{traceeven} and \eqref{harmoniceven1} to eliminate $H_0$ and $H_2$ as defined in Eq.~\eqref{evenpart}. Then, we use a generalization of the Berndtson-Zerilli transformations:
\begin{align}
&\frac{H_1}{2}=\left[\frac{i\omega(M-r)}{fr^3}+\mu ^2\frac{3 i r  \omega }{2 M+r^3 \mu ^2}\right]\varphi_0+\frac{i\omega}{r}\frac{d\varphi_0}{dr}\,,\nonumber\\
&\frac{K}{2}=\left[\frac{f}{r^3}-\mu ^2\frac{ 6 r+r^3 \mu ^2-10 M}{2\left(2 M r+r^4 \mu ^2\right)}\right]\varphi_0-\frac{f}{r^2}\frac{d\varphi_0}{dr}\,.\nn
\end{align}
After substituting these transformations into the system of equations we arrive at a single wave equation of the form: 
\begin{equation}
\label{evenl0}
\frac{d^2}{dr_*^2} \varphi_0 + \left[\omega^2-V_0(r)\right]\varphi_0=0\,,
\end{equation}
with
\begin{equation}
V_0=f\left[\frac{2M}{r^3}+\mu^2+\frac{24M(M-r)\mu^2+6 r^3 (r-4 M) \mu ^4}{\left(2 M+r^3 \mu ^2\right)^2}\right]\,.\nn
\end{equation}
In this form it is clear that in the massless limit the monopole reduces to the scalar-field wave equation with $l=0$ \cite{Berti:2009kk}.

We have solved Eq.~\eqref{evenl0} subjected to appropriate boundary conditions (regularity at the horizon and at infinity, see also next sections) by direct integration,
looking for eigenvalues $\omega=\omega_R+i\omega_I$. Given the time dependence \eqref{decom}, stable modes are characterized by $\omega_I<0$
and unstable modes by $\omega_I>0$. We found one unstable mode, detailed in Fig.~\ref{fig:GL} and characterized by a purely imaginary, positive
component. This is a low-mass instability which disappears for $M\mu\geq 0.43$ and has a minimum growth timescale of around $M\omega_I\sim 0.046$.
In fact, as recognized very recently~\cite{Babichev:2013} while our own work was in its final stages, the linearized equations~\eqref{eqmotioncurved} are equivalent to those describing four-dimensional perturbations of a five-dimensional black string after a Kaluza-Klein reduction of the extra dimension. Therefore, the system is affected by Gregory-Laflamme instability~\cite{Gregory:1993vy,Kudoh:2006bp} that manifests itself in the spherically symmetric, monopole mode.
One interesting aspect of our own formulation is that we are able to reduce this instability to the study of a very simple wave equation,
described by~\eqref{evenl0}.

To summarize, in this setup Schwarzschild BHs are unstable. The instability timescale depends strongly on the mass scale $\mu$.
For low masses, we find numerically that $\omega_I\sim 0.7\mu$, in good agreement with analytic calculation by Camps and Emparan~\cite{Camps:2010br}. 

The Gregory-Laflamme instability only affects spherically-symmetric ($l=0$) modes~\cite{Kudoh:2006bp}, so we expect the rest of the sector to be stable. We confirm this result in Sec.~\ref{sec:schwar} below, where we derive the complete linear dynamics on a Schwarzschild metric. 

A more relevant question is related to the role of a cosmological constant. When the background metrics are two copies of Schwarzschild-de Sitter solutions, the field equations~\eqref{eqmotioncurved} do not arise from a Kaluza-Klein decomposition of a five-dimensional black string. Thus, it is not obvious a priori if the monopole instability discussed above survives when $\Lambda_g=\Lambda_f\neq0$. 

Our formalism can be immediately extended to accommodate Schwarzschild-de Sitter backgrounds. In this case, Eq.~\eqref{eq_m} is modified with new terms proportional to $\Lambda_g$, see e.g. Eq.~(2.1) in Ref.~\cite{Hassan:2012gz}. From the latter equation, one obtains the same divergenceless and traceless conditions as in Eqs.~\eqref{constraint1} and~\eqref{constraint2}. Finally, using these conditions and the commutator of two covariant derivatives, it turns out that the linearized field equation is precisely as in Eq.~\eqref{eqmotioncurved}. That is, terms that explicitly depend on $\Lambda_g$ cancel out and the only contribution of the cosmological constant arises through background quantities. From the system~\eqref{eqmotioncurved}--\eqref{constraint2}, it is straightforward to obtain a master equation for spherical perturbations of Schwarzschild-de Sitter BHs. Here we omit the details and only give the final result. The monopole is described by an equation of the same form as Eq.~\eqref{evenl0}, but where the potential now reads:
\begin{eqnarray}
 &&V_0^{\Lambda_g}=\frac{1-2M/r-\Lambda_g /3\,r^2}{r^3 \left[2 M+r^3 \left(\mu ^2-2 {\Lambda_g/3}\right)\right]^2}\nn\\
 &&\times\left\{8 M^3+12 M^2 r^3 \left(3 \mu ^2-8 {\Lambda_g/3}\right)\right.\nn\\
 &&\left.+r^7 \left(\mu ^2-2 {\Lambda_g/3}\right)^2 \left[6+r^2 \left(\mu ^2-2 {\Lambda_g/3}\right)\right]\nn\right.\\
 &&\left.-6 M r^4 \left(\mu ^2-2 {\Lambda_g/3}\right) \left[4+r^2 \left(3 \mu ^2-10 {\Lambda_g/3}\right)\right]\right\}\,.\label{V0dS}
\end{eqnarray}
Using the same technique as before, we have integrated Eq.~\eqref{evenl0} with the potential~\eqref{V0dS}. The results are shown in Fig.~\ref{fig:GL} for various values of $\Lambda_g=\Lambda_f$. Note that massive spin-2 perturbations propagating in an asymptotically de Sitter spacetime are subjected to the bound $\mu^2>2\Lambda_g/3$~\cite{Higuchi1987397}. Below such bound, the helicity-0 component of the massive graviton becomes a ghost.
When the bound is saturated, $\mu^2=2\Lambda_g/3$, the helicity-0 mode becomes pure gauge and the instability disappears. Theories with such fine-tuning are called ``partially massless gravities''~\cite{Deser:1983mm,Deser:2001pe} [see also Refs.~\cite{Hassan:2012gz,Deser:2012qg,Deser:2013uy,Hassan:2012rq,Hassan:2013pca}] and they are not affected by the monopole instability discussed above.
Finally, as shown in Fig.~\ref{fig:GL}, the instability is even more effective for Schwarzschild-de Sitter BHs and it exists roughly in the same range of graviton mass.


For both Schwarzschild and Schwarzschild-de Sitter BHs, the instability timescale is of the order of the Hubble time when $\mu\sim 2\times 10^{-33} {\rm eV}$~\cite{Babichev:2013}.
This of course, does {\it not} mean that the observation of compact objects imposes constraints on the graviton mass \footnote{The monopole instability does not impose limits on the
graviton mass, but the observation of rotating compact BHs, discussed later on, does impose strict limits on the graviton mass.}.
Rather, it suggests that the background solution used to describe these geometries is likely not the physical one.
It would seem that a suitable background geometry is given by the end-state of this monopole instability.

Our linear analysis cannot handle the nonlinear development of the instability, nor the nonlinear final state.
However, from the mode profile in Fig.~\ref{fig:GL}, it is tempting to conjecture that a Schwarzschild BH surrounded by a graviton
cloud could be a possible solution of the field equations. We note that this endstate is completely different, as it must be,
from the standard Gregory-Laflamme instability which acts to fragment black strings \cite{Cardoso:2006ks,Lehner:2010pn}.

\section{Massive spin-2 fields on a Schwarzschild background}\label{sec:schwar}
We have established the instability of spherically symmetric fluctuations in 
non-rotating backgrounds. We now generalize the analysis to the full set of non-axisymmetric polar and axial perturbations.

\subsection{Axial sector} 
The axial field equations are derived in Appendix~\ref{app:decomposition}.
The axial sector is fully described by the following system:
\beq
&&\frac{d^2}{dr_*^2}Q+\left[\omega^2-f\left(\mu^2+\frac{\Lambda+4}{r^2}-\frac{16M}{r^3}\right)\right]Q=S_Q\,,\label{oddf1}\\
&&\frac{d^2}{dr_*^2}Z+\left[\omega^2-f\left(\mu^2+\frac{\Lambda-2}{r^2}+\frac{2M}{r^3}\right)\right]Z=S_Z\,,\label{oddf2}
\eeq
where $\Lambda=l(l+1)$ and we have defined the tortoise coordinate $r_*$ via $dr/dr_*=f\equiv1-2M/r$. The functions $Q(r)\equiv f(r)h_1$ and $Z(r)\equiv h_2/r$ are combinations of the axial perturbations as defined in Eq.~\eqref{oddpart}, whereas the source terms are given by
\beq
S_Q&=& (\Lambda-2)\frac{2f(r-3M)}{r^3}Z\,,\\
S_Z&=& \frac{2}{r^2}f\,Q\,.
\eeq
%
\subsubsection{Axial dipole mode}
The $l=0$ monopole mode does not exist in the axial sector since the angular part of the axial perturbations \eqref{oddpart} vanishes for $l=0$. For the dipole mode ($l=1$ or equivalently $\Lambda=2$), the angular functions $W_{lm}$ and $X_{lm}$ vanish and one is left with a single decoupled equation:
\begin{equation}
\label{oddl1}
\frac{d^2}{dr_*^2} Q + \left[\omega^2-f\left(\mu^2+\frac{6}{r^2}-\frac{16M}{r^3}\right)\right]Q=0\,.
\end{equation}
%
\subsubsection{Axial massless limit}
It is interesting to note that in the massless limit we can use the transformations
\begin{align}
&h_0=\frac{1}{i\omega}\left[\varphi_1+\frac{\Lambda-2}{3}\varphi_2\right]\,,\nonumber\\
&h_1=\frac{1}{(i\omega)^2}\left[\frac{2}{r}\varphi_1+\frac{2-\Lambda}{3r}\varphi_2-\frac{d\varphi_1}{dr}+\frac{2-\Lambda}{3}\frac{d\varphi_2}{dr}\right]\,,\nonumber\\
&h_2=\frac{1}{(i\omega)^2}\left[\varphi_1+\frac{(\Lambda+1)r-6M}{3r}\varphi_2+(r-2M)\frac{d\varphi_2}{dr}\right]\,,\nn
\end{align}
to reduce the system to a pair of decoupled equations, given by a ``vectorial'' and a ``tensorial'' Regge-Wheeler equation
\be
\frac{d^2}{dr_*^2} \varphi_s + \left[\omega^2-f\left(\frac{\Lambda}{r^2}+(1-s^2)\frac{2M}{r^3}\right)\right]\varphi_s=0\,,\label{RW}
\ee
where $s=0,1,2$ for scalar, vectorial, or tensorial perturbations. These transformations were first found by Berndtson~\cite{Berndtson:2009hp} when studying the massless graviton perturbations of the Schwarzschild metric in the harmonic gauge. In the massless limit the vectorial degree of freedom can be removed by a gauge transformation, but for $\mu\neq 0$ it becomes a physical mode.
Note that the wave equation \eqref{RW} for $s=1$ is identical to that describing electromagnetic perturbations of Schwarzschild BHs \cite{Berti:2009kk}; thus the axial spectrum of massive spin-2 perturbations should include a mode which approaches that of an electromagnetic mode in the low-mass limit.

\subsection{Polar sector} 
The polar equations are more involved and derived in Appendix~\eqref{app:decomposition}. The polar sector is fully described by a system of three coupled ordinary differential equations:
\beq
\label{polar_eq1}
f^2\frac{d^2 K}{dr^2}+\hat\alpha_1 \frac{d K}{dr}+\hat\beta_1 K &=& S_K\,,\\
\label{polar_eq2}
f^2\frac{d^2 \eta_1}{dr^2}+\hat\alpha_2 \frac{d \eta_1}{dr}+\hat\beta_2 \eta_1 &=& S_{\eta_1}\,,\\
\label{polar_eq3}
f^2\frac{d^2 G}{dr^2}+\hat\alpha_3 \frac{d G}{dr}+\hat\beta_3 G &=& S_G\,,
\eeq
where the dynamical variables $K$, $\eta_1$ and $G$ are defined in Eq.~\eqref{evenpart} and the source terms are given by
\begin{align}
\label{source1}
 S_K&= \Lambda\hat\gamma_1 \frac{d \eta_1}{dr}+\hat\delta_1\Lambda \eta_1+\Lambda(\Lambda-2)\hat\sigma_1 \frac{d G}{dr}+\Lambda(\Lambda-2)\hat\rho_1 G\,,\\
\label{source2} 
S_{\eta_1}&= \hat\gamma_2 \frac{d K}{dr}+\hat\delta_2 K+\Lambda(\Lambda-2)\hat\sigma_2 \frac{d G}{dr}+\Lambda(\Lambda-2)\hat\rho_2 G\,,\\
\label{source3}
 S_G &= \hat\gamma_3 \frac{d K}{dr}+\hat\delta_3 K+\hat\sigma_3 \frac{d \eta_1}{dr}+\hat\rho_3 \eta_1\,.
\end{align}
The coefficients $\hat\alpha_i,\,\hat\beta_i,\,\hat\gamma_i,\,\hat\delta_i,\,\hat\sigma_i,\,\hat\rho_i$ are radial functions which also depend on $\omega$ and $l$. These equations are rather lengthy and since their explicit form is not fundamental here, we made them available online in {\scshape Mathematica} notebooks~\cite{webpage}.


\subsubsection{Polar dipole mode}
The polar monopole was already investigated in Section~\ref{sec:monopoleinstability} and shown to lead to Gregory-Laflamme-like instabilities~\cite{Babichev:2013}.
We now study the dipole mode. In the dipole case, $l=1$, $\Lambda=2$, the radial function $G$ identically vanishes and we are left with a pair of coupled equations satisfying the following system:
\beq
\label{polar_dipole1}
f^2\frac{d^2 K}{dr^2}+\hat\alpha_1 \frac{d K}{dr}+\hat\beta_1 K &=& 2(\hat\gamma_1 \frac{d \eta_1}{dr}+\hat\delta_1 \eta_1)\,,\\
\label{polar_dipole2}
f^2\frac{d^2 \eta_1}{dr^2}+\hat\alpha_2 \frac{d \eta_1}{dr}+\hat\beta_2 \eta_1 &=& \hat\gamma_2 \frac{d K}{dr}+\hat\delta_2 K\,.
\eeq
%
\subsubsection{Polar massless limit}
In the massless limit we can use the argument presented by Berndtson in Ref.~\cite{Berndtson:2009hp} to reduce the system to three decoupled equations, one ``scalar'', one ``vectorial''~\eqref{RW} and one ``tensorial'' equation described by Zerilli's equation~\cite{Zerilli:1970se}
\footnote{Note that in these transformations there are four functions. One tensorial, one vectorial, and two scalars. However one of the scalar functions is simply the trace of $h_{\mu\nu}$, which vanishes in our case (in their notation is the scalar function $\varphi_0$, not to be confused with the scalar function used here). We stress again the importance of having a vanishing trace in order to have a correct number of degrees of freedom.}.
In the massless limit the scalar and the vectorial degrees of freedom can be removed by a gauge transformation but, for $\mu\neq 0$, they become physical. Thus, we expect that the small-mass limit of massive gravity spectrum includes a family of modes which
are identical to that of a scalar and an electromagnetic mode (these modes are discussed in Ref.~\cite{Berti:2009kk} and available online at 
\cite{webpage}).

\subsection{Results}
We have solved the previous systems of equations subjected to appropriate boundary conditions,
which defines an eigenvalue problem for the complex frequency $\omega\equiv\omega_R+i\omega_I$; this problem can be solved using 
several different techniques~\cite{Berti:2009kk,PaniNRHEP2} which we detail in Appendix \ref{app:modes}.

In general, the asymptotic behavior of the solution at infinity is given by
\be
\Phi_j(r)\sim B_j e^{-i k_{\infty} r}r^{-\frac{M(\mu^2-2\omega^2)}{k_{\infty}}}+C_j e^{i k_{\infty} r}r^{\frac{M(\mu^2-2\omega^2)}{k_{\infty}}}\,,\nonumber
\ee
where $k_{\infty}=\sqrt{\mu^2-\omega^2}$ and, without loss of generality, we assume Re$(k_{\infty})>0$. The spectrum of massive perturbations admits two different families of physically motivated modes, which are distinguished according to how they behave at spatial infinity. The first family includes the standard QNMs, which corresponds to purely outgoing waves at infinity, i.e., they are defined by $B_j=0$~\cite{Berti:2009kk}. The second family includes quasibound states, defined by $C_j=0$. The latter correspond to modes spatially localized within the vicinity of the BH and that decay exponentially at spatial infinity~\cite{Dolan:2007mj,Rosa:2011my,Pani:2012bp,PaniNRHEP2}.

\subsubsection{Quasinormal modes}
%
\begin{figure}[htb]
\begin{center}
\epsfig{file=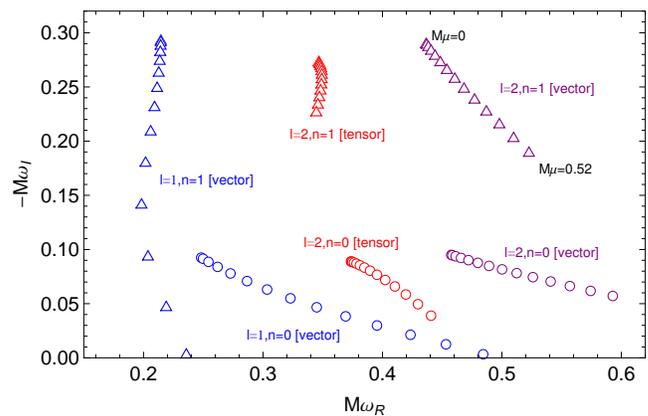,width=8.5cm,angle=0,clip=true}
\caption{QNM frequencies for axial $l=1,2$ modes, for a range of field masses $M\mu=0,0.04,\ldots,0.52$. Points with largest $|\omega_I|$ correspond to $\mu\to0$. The fundamental mode ($n=0$, circles) and the first overtones ($n=1$, triangles) are shown. In the massless limit the ``vector'' modes have the same QNM frequency as the electromagnetic field, and the ``tensor'' modes have the same QNM frequency as the massless gravity perturbations. \label{fig:QNM}}
\end{center}
\end{figure}
The axial QNM frequencies for different values of the spin-2 mass are shown in Figure~\ref{fig:QNM}. 
As expected, for $l\geq 2$ one can sensibly group the modes in two families for any given $l$ and $n$. They can be distinguished by their behavior in the massless limit, the spectrum of the ``vector'' modes reduces to the spectrum of the photon, while the ``tensor'' modes, which are the only physical modes in the massless limit, approaches the spectrum of the massless gravity perturbations.   
For the lowest overtones, as the mass increases the decay rate decreases to zero, reaching a limit where the QNM disappears. This is linked with the decreasing height of the effective potential barrier as was previously discussed in Ref.~\cite{Ohashi:2004wr}. The limiting behavior, when the damping rate reaches zero are the so-called \emph{quasiresonant} modes, which were already shown to occur for massive scalar~\cite{Ohashi:2004wr,Konoplya:2004wg} and massive vector~\cite{Konoplya:2005hr} fields.     

Polar QNMs are more challenging to compute, because the perturbation equations are lengthy and translate into higher-term recurrence relations in a matrix-valued continued-fraction method~\cite{PaniNRHEP2}. On the other hand, due to the well-known divergent nature of the QNM eigenfunctions~\cite{Berti:2009kk}, a direct integration is not well suited to compute these modes precisely. Instead of computing these modes, in the following we shall rather focus on quasibound states --~both in the axial and polar sector~-- which are easier to compute [cf. Appendix~\ref{app:modes}] and more relevant for our discussion.

\subsubsection{Quasibound states}

Besides the QNM spectrum, massive fields can also be localized in the vicinity of the BH, showing a rich spectrum of quasibound states with complex frequencies. Here the terminology `quasi' stands for the fact that these states decay due to the absorption by the BH, hence the complex frequencies. Bound states were already considered for massive scalar~\cite{Dolan:2007mj}, Dirac~\cite{Gal'tsov:1983wq,Lasenby:2002mc} and Proca~\cite{Gal'tsov:1984nb,Rosa:2011my} fields. In the small-mass limit $M\mu\ll l$, it was shown that for these fields the spectrum resembles that of the hydrogen atom: 
\be
\omega_R/\mu \sim 1-\frac{(M\mu)^2}{2(j+1+n)^2}\,, \label{hydrogenic}
\ee
where $j=l+S$ is the total angular momentum of the state with spin projections $S=-s,-s+1,\ldots,s-1,s$. Here $s$ is the spin of the field. For a given $l$ and $n$, the total angular momentum $j$ satisfies the quantum mechanical rules for addition of angular momenta, $|l-s|\leq j\leq l+s$.
\begin{figure*}[htb]
\begin{center}
\begin{tabular}{cc}
\epsfig{file=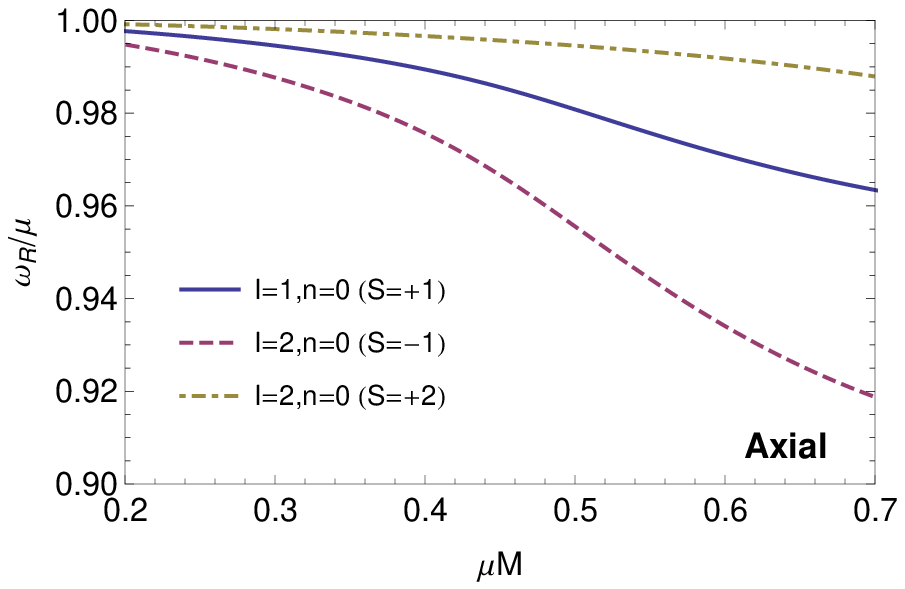,width=7.5cm,angle=0,clip=true}
\epsfig{file=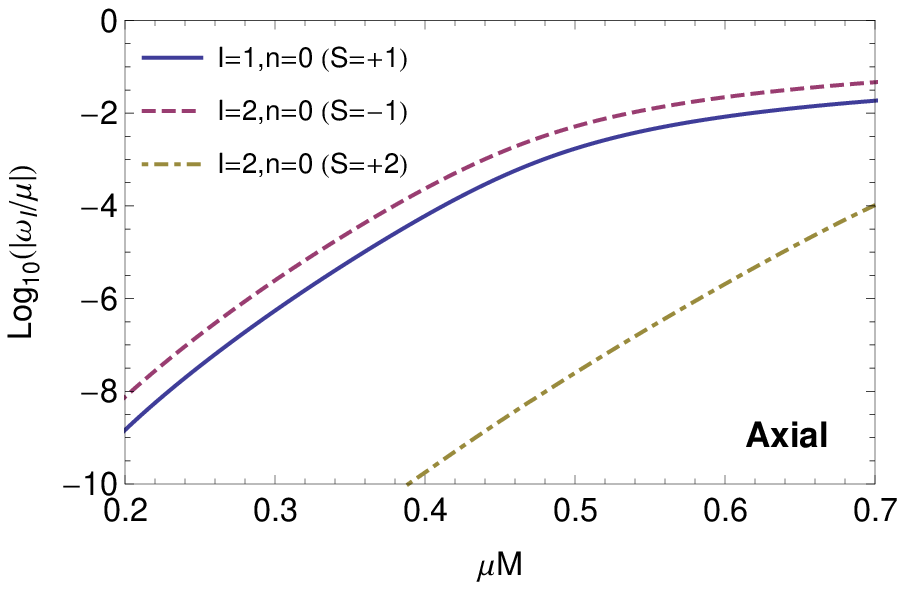,width=7.5cm,angle=0,clip=true}\\
\epsfig{file=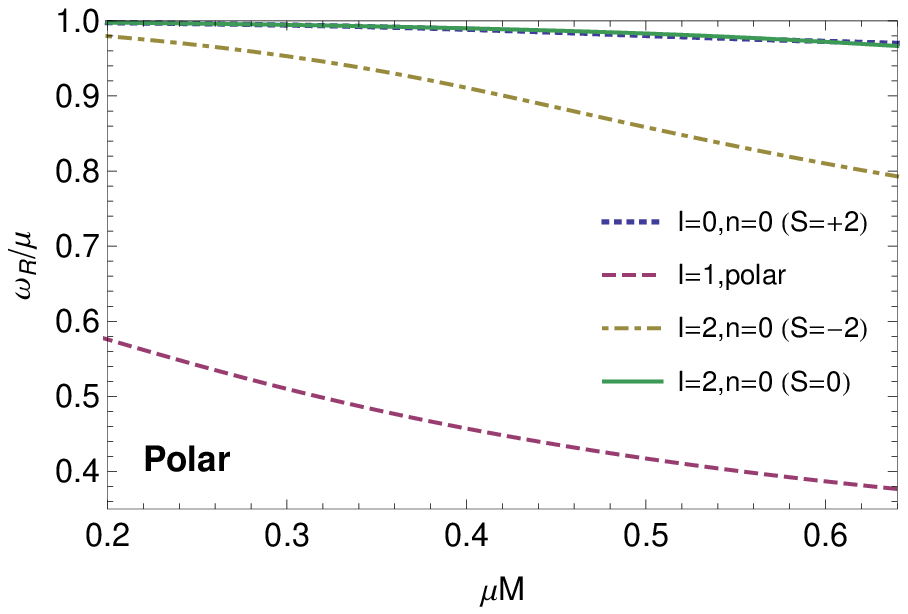,width=7.5cm,angle=0,clip=true}
\epsfig{file=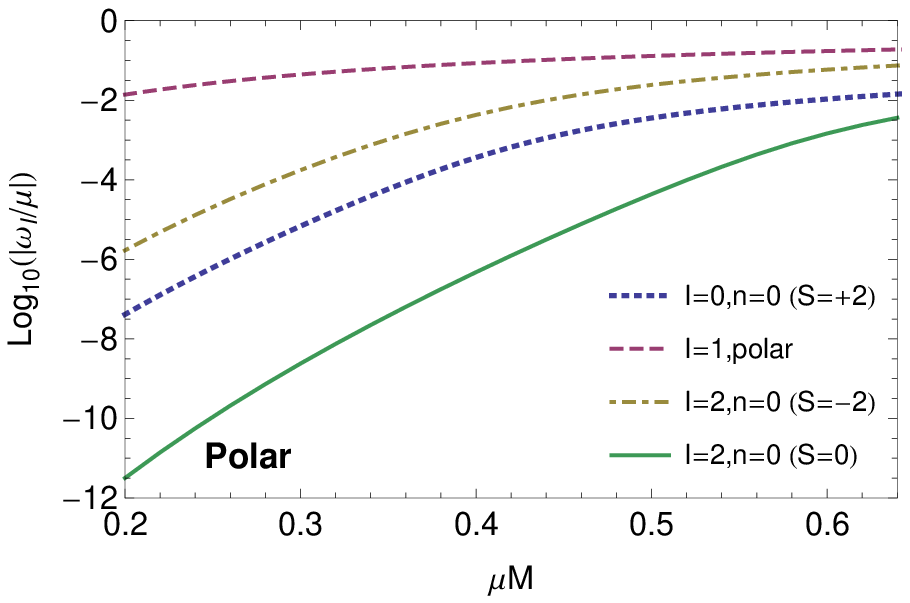,width=7.5cm,angle=0,clip=true}
\end{tabular}
\caption{Axial (Top) and polar (bottom) quasibound state levels of the massive spin-2 field. The left and right panels show the real part, $\omega_R/\mu$, and the imaginary part, $\omega_I/\mu$, of the mode as a function of the mass coupling $M\mu$, respectively. We label the modes by their angular momentum $l$, overtone number $n$ and spin projection $S$. Except for the polar dipole $l=1$, the spectrum is hydrogenic in the massless limit. \label{fig:BS}}
\end{center}
\end{figure*}

Our results show that the spectrum \eqref{hydrogenic} also describes massive spin-2 perturbations which is also confirmed analytically for the axial mode $l=1$ (see Eq.~\eqref{ana_real} of Appendix~\ref{app:ana}).
In Fig.~\ref{fig:BS} we show the quasibound-state frequency spectrum for the lowest modes. Apart from the polar dipole (we discuss this in detail below), all other modes follow a hydrogenic spectrum as predicted by Eq.~\eqref{hydrogenic}. The monopole $l=0$ [which belongs to a different family than the unstable monopole mode discussed in Sec.~\ref{sec:monopoleinstability}] is fully consistent with $S=+2$ which is in agreement with the rules for the sum of angular momenta, $|l-s|\leq j\leq l+s\implies j=2$. For each pair $l\geq 2$ and $n$ there are five kinds of modes, characterized by their spin projections. Here we do not show the mode $l=2$, $n=0$, $S=1$, which is very difficult to find numerically due the complicated form of the polar equations and his tiny imaginary part. Besides that, the existence of the mode $l=2$, $n=1$, $S=0$ with approximately the same real frequency makes it even more challenging to evaluate the $l=2$, $n=0$, $S=1$ mode with sufficient precision.  

\begin{figure}[htb]
\begin{center}
\epsfig{file=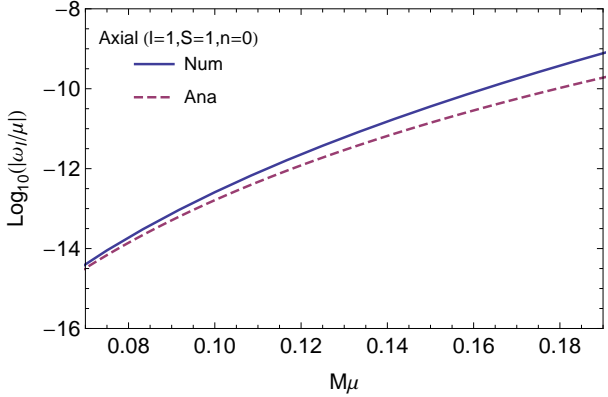,width=8cm,angle=0,clip=true}
\caption{Comparison between the numerical and analytical results for the the axial mode $l=1,\,S=1,\,n=0$ as a function of the mass coupling $M\mu$. The solid line shows the numerical data and the dashed shows the analytical formula~\eqref{wI_ana}.\label{fig:axial_ana}}
\end{center}
\end{figure}

Evaluating the dependence of $\omega_I(\mu)$ in the small-$M\mu$ limit turns out to be extremely challenging, due to the fact that
$\omega_I$ is extremely small in this regime. Our results indicate a power-law dependence of the kind found previously for other massive fields~\cite{Rosa:2011my}, $\omega_I/\mu\propto -(M\mu)^{\eta}$, with
\be
\eta=4l+2S+5\,. \label{wIslope}
\ee
The fact that the modes $l=L$, $S=S_1$ and $l=L+S_1$, $S=-S_1$ have the same exponent is a further confirmation of this scaling. Note that only the constant of proportionality depends on the overtone number $n$ and it also generically depends on $l$ and $S$. This is confirmed analytically for the axial mode $l=1,\,S=1\,,n=0$, as shown in Fig.~\ref{fig:axial_ana}, where we see that in the low-mass limit the numerical results approaches the analytical formula derived in Appendix~\ref{app:ana}, given by
\be
\omega_I/\mu\approx -\frac{320}{19683}(M\mu)^{11}\,. \label{wI_ana}
\ee
%

The quasibound state found for the polar dipole is clearly the more interesting. This mode appears to be isolated from the rest of the modes and it does not follow the small-mass behavior predicted by Eqs.~\eqref{hydrogenic} and~\eqref{wIslope}. Furthermore, we have found only a single fundamental mode for this state, and no overtones. For this mode, the real part is much smaller than the mass of the spin-2 field. 

The real part of this special mode in region $M\mu\lesssim 0.4$ is very well fitted by
\be
\label{polar_di_Re}
\omega_R/\mu\approx 0.72(1-M\mu)\,.
\ee
For the imaginary part we find in the limit $M\mu\ll 1$,
\be
\omega_I/\mu\approx -(M\mu)^{3}\,.
\ee
That this mode is different is not completely unexpected since in the massless limit it becomes unphysical. This peculiar behavior seems to be the 
result of a nontrivial coupling between the states with spin projection $S=-1$ and $S=0$. Besides that, this mode has the largest binding energy ($\omega_R/\mu-1$) for all couplings $M\mu$, much higher than the ground states of the scalar, Dirac and vector fields (see Fig.7 of Ref.~\cite{Rosa:2011my}). However the decay rate is very large even for small couplings $M\mu$, corresponding to a very short lifetime for this state.

To summarize, the $l>0$ modes of Schwarzschild BHs in massive gravity theories are stable, with a rich and potentially interesting 
fluctuation spectrum, which could give rise to very long-lived clouds of tensor hair in the right circumstances.
We now show that once rotation is included, this hair grows exponentially and extracts angular momentum away from the BH. Thus, while the monopole $l=0$ mode is unstable even in the static case, the $l>0$ modes suffer for a superradiant instability only above a certain threshold of the BH angular momentum.

\section{Massive spin-2 perturbations of slowly rotating Kerr BHs}\label{sec:kerr}
In Ref.~\cite{Pani:2012bp} a method to study generic perturbations of slowly rotating BHs was developed. Here we extend this method to massive spin-2 perturbations of slowly rotating Kerr BHs. We derive the linearized field equations to first order in $\tilde{a}$, although our analysis can be generalized to higher order in the BH angular momentum. 

The technique is detailed in Appendix~\ref{app:kerr} and it consists in a decomposition of the perturbation equations in tensor spherical harmonics and in a expansion in the BH angular momentum. The method was originally developed to study the gravitational perturbations of slowly-rotating stars~\cite{Kojima:1992ie,1993ApJ...414..247K,1993PThPh..90..977K} and it has been recently applied to BH spacetimes~\cite{Pani:2012vp,Pani:2013ija}. As a result of using a basis of spherical harmonics in a nonspherical background, the perturbation equations display parity-mixing and coupling among perturbations with different harmonic indices. However, as discussed in Ref.~\cite{Pani:2012vp}, to first order in $\tilde{a}$ the eigenvalue spectrum is described by two decoupled sets, one for the axial and one for the polar perturbations, and all harmonic indices decoupled. In the following we discuss the axial and polar sector separately.

\subsection{Axial equations at first order}
The field equations are derived in Appendix~\ref{app:kerr}, where the method to separate the equations is shown. 
By defining:
\begin{eqnarray}
 h_1(r)&=& \frac{Q(r)}{f(r)}\left(1-\frac{\tilde{a}m M^2 \left(\Lambda+2\right) }{\Lambda r^3 \omega }\right)\,,\\
 h_2(r)&=& Z(r)r\left(1-\frac{\tilde{a}m M^2 \left(\Lambda-2\right) }{\Lambda r^3 \omega }\right)\,,
\end{eqnarray}
we obtain that a fully consistent solution at first order is such that $Z$ and $Q$ satisfy the following equations:
\begin{eqnarray}\label{axial1}
 \frac{d^2Q}{dr_*^2}+V_Q Q(r)&=&S_Q Z(r)\,,\\
\label{axial2}
 \frac{d^2Z}{dr_*^2}+V_Z Z(r)&=&S_Z Q(r)
\end{eqnarray}
with
\begin{align}
&V_Q=\omega^2-\frac{4 \tilde{a}m M^2\omega }{r^3}-\nn\\
&f\left[\frac{\Lambda+4}{r^2}-\frac{16 M}{r^3}+\mu ^2+\tilde{a}m M^2\frac{6 (4 r-9 M)(\Lambda+2)}{\Lambda r^6 \omega }\right]\,,\\
& V_Z=\omega^2-\frac{4 \tilde{a}m M^2\omega }{r^3}-\nn\\
&f\left[\frac{\Lambda-2}{r^2}+\frac{2M}{r^3}+\mu ^2+\tilde{a}m M^2\frac{6 (\Lambda -2)  (r-3M)}{\Lambda  r^6 \omega }\right]\,,\\
&S_Q= 2(\Lambda-2)f\left[\frac{r-3M}{r^3}\right.\nn\\
&\left.-\tilde{a}m M^2\frac{\left(6 M (4+\Lambda )-r \left(10+3 \Lambda +3 r^2 \omega ^2\right)\right)}{\Lambda r^6 \omega }\right]\,,\\
&S_Z=2f\left[\frac{1}{r^2}+\tilde{a}m M^2\frac{\left(-10+3 \Lambda +3 r^2 \mu ^2\right)}{\Lambda  r^5 \omega }\right]\,.
\end{align}
These equations reduce to Eqs.~\eqref{oddf1} and~\eqref{oddf2} in the nonrotating limit.
In the dipole case $l=1$, $\Lambda=2$, the function $Z$ vanishes and we are left with a single decoupled equation:
\begin{equation}\label{axialdi_kerr}
 \frac{d^2Q}{dr_*^2}+V_Q Q(r)=0\,.
\end{equation}

\subsection{Polar equations at first order}
In line with the non-rotating case, for the polar sector we obtain at first order in $\tilde{a}$ three coupled equations for $K$, $\eta_1$ and $G$, which generalize Eqs.~\eqref{polar_eq1},~\eqref{polar_eq2}, and~\eqref{polar_eq3}, but in this case the coefficients $\hat\alpha_i,\,\hat\beta_i,\,\hat\gamma_i,\,\hat\delta_i,\,\hat\sigma_i,\,\hat\rho_i$ are also functions of $m\tilde{a}$. Due to the length of the equations we do not show them explicitly here but we made them available online in {\scshape Mathematica} notebooks~\cite{webpage}.

\subsection{Superradiance and quasibound states}
Interesting phenomena, such as BH superradiance, are already manifest at first order in the BH angular momentum. A second order approximation would be necessary to consistently describe superradiance (see e.g. Ref.~\cite{Pani:2012bp}) but this is beyond the scope of this work. 

As for the Schwarzschild case, at the horizon we must impose regular boundary conditions, which correspond to purely ingoing waves,
\be
\label{BC_hor_Kerr}
\Phi_j(r)\sim e^{-ik_H r_*}\,,
\ee
as $r_*\to -\infty$, where
\be
k_H=\omega-m\Omega_H=\omega-\frac{m\tilde{a}}{4M}+\mathcal{O}(\tilde{a}^3)\,. 
\ee
Here the horizon angular velocity $\Omega_H=a/(2Mr_+)$ was expanded to first-order in rotation. When $k_H<0$ an observer at infinity will see waves emerging from the BH~\cite{Teukolsky:1973ha}. This corresponds to the superradiant condition $\omega<m\Omega_H$~\cite{Teukolsky:1974yv}, which at first-order in the rotation amounts to
\be
\tilde{a}>\frac{4M\omega_R}{m}\,,
\ee
where $\omega_R$ is the real part of the mode frequency, $\omega=\omega_R+i\omega_I$. All the polar and axial equations can be brought to a form such that the near-horizon solution is given by Eq.~\eqref{BC_hor_Kerr}. We thus expect that superradiance will also occur for massive spin-2 fields even at first-order in the rotation.

Superradiant scattering leads to instabilities of bosonic massive fields~\cite{Detweiler:1980uk,Dolan:2007mj,Pani:2012bp,Witek:2012tr,Dolan:2012yt}. This instability was explicitly shown for scalars and vectors, but generic arguments indicate that it is present for other integer-spin fields.
Note that with our convention, unstable modes correspond to $\omega_I>0$. These superradiant instabilities occur only for waves localized in the vicinity of the BH, i.e., quasibound states, so we focus on these states in the next sections. 

The continued fraction method can be used to determine the quasibound state frequencies of the axial equations by imposing an appropriate \emph{ansatz} which in this case is given by
\be
\label{ansatz2}
\Phi_j(\omega,r)=f(r)^{-2ik_H}r^{\nu}e^{-qr}\sum_n{a^{(j)}_n}f(r)^n\,,
\ee
where $\nu=-q+\omega^2/q$. To compute the quasinormal mode frequencies we use $q=-\sqrt{\mu^2-\omega^2}$ and for the quasibound state frequencies $q=\sqrt{\mu^2-\omega^2}$.
Inserting Eq.~\eqref{ansatz2} into Eq.\eqref{axialdi_kerr} leads to a six-term recurrence relation which can be reduced to a three-term recurrence relation by successive Gaussian elimination steps~\cite{Leaver:1990zz,Onozawa:1995vu}. For $l\geq 2$ we find a six-term matrix-valued recurrence relation which can also be brought to a three-term recurrence relation using a matrix-valued Gaussian elimination. The explicit form of the coefficients is not shown here for brevity but it is available online~\cite{webpage}.

Although the continued-fraction method works very well for quasibound states, the multiple matrix inversion of almost singular matrices (since some matrices are proportional to $\tilde{a}$) makes it very difficult to compute the very small imaginary part of the axial quasibound states. We therefore use the direct integration method for both the polar and axial quasibound states which gives more accurate results in this case, and use the continued-fraction method to check the robustness of our results.

\subsection{Results}
\begin{figure*}[htb]
\begin{center}
\epsfig{file=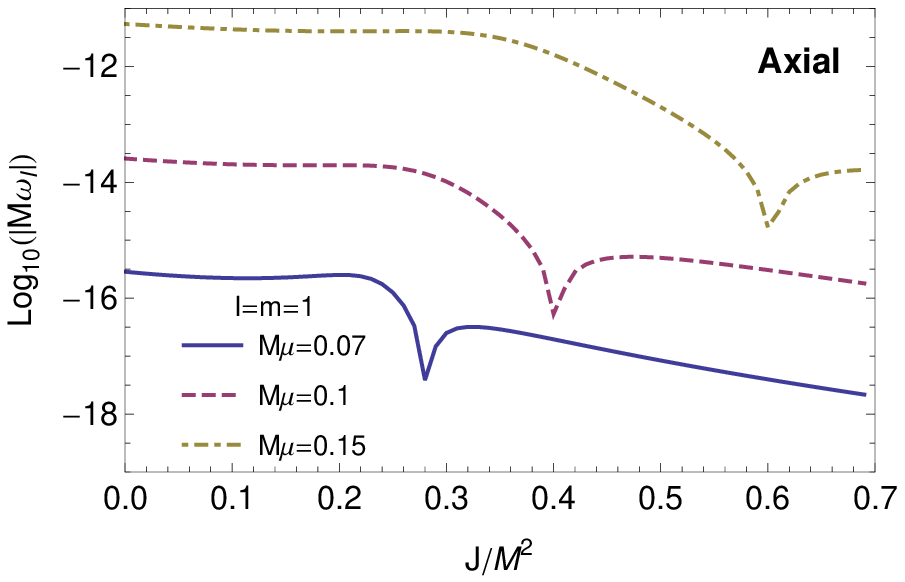,width=7cm,angle=0,clip=true}
\epsfig{file=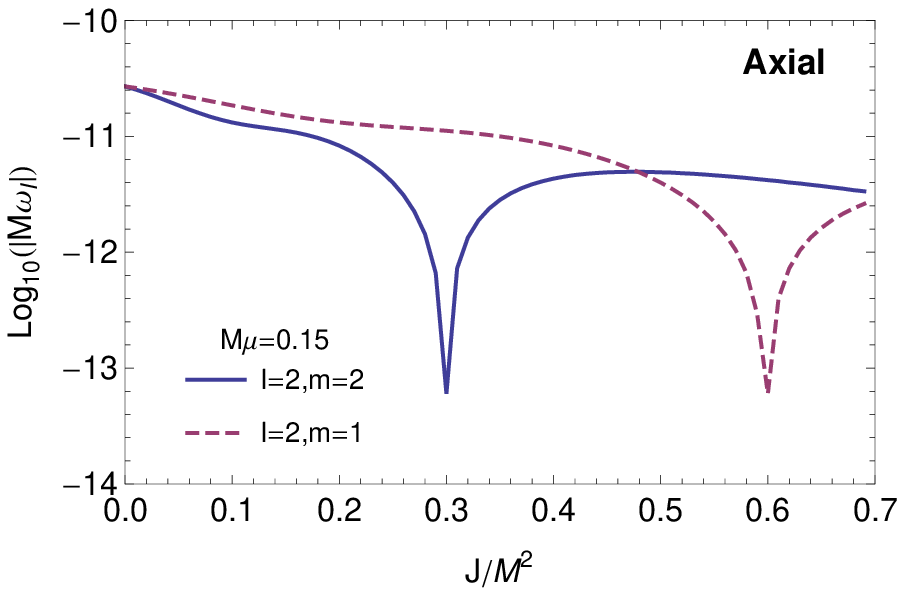,width=7cm,angle=0,clip=true}\\
\epsfig{file=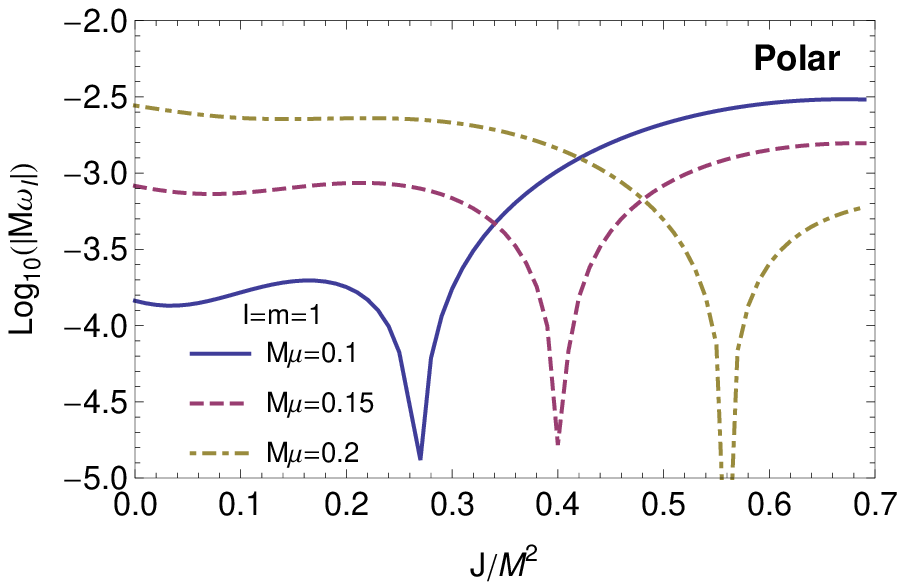,width=7cm,angle=0,clip=true}
\epsfig{file=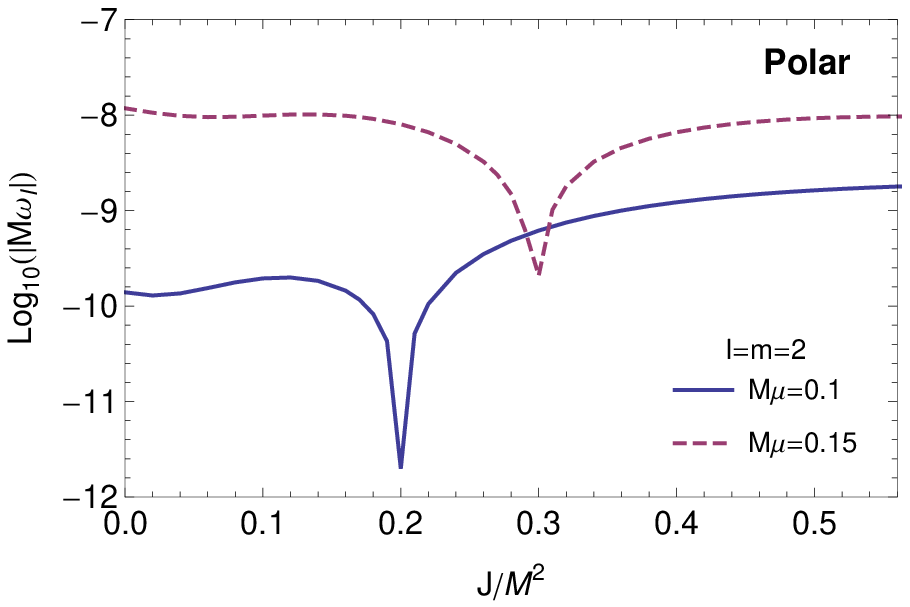,width=7cm,angle=0,clip=true}
\caption{Absolute value of the imaginary part of the axial and polar quasibound modes as a function of the BH rotation rate $\tilde{a}$ for different values of $l$ and $m$ and different values of the mass coupling $\mu M$, computed at first order. Left top panel: axial dipole for $l=m=1$. Right top panel: axial mode $S=-1$ for a mass coupling $M\mu=0.15$ and different values of $m$.
Left bottom panel: polar dipole mode for $l=m=1$. Right bottom panel: polar mode $l=m=2$, $S=-2$. 
For any mode with $m\geq 0$, the imaginary part crosses the axis and become unstable when the superradiance condition is met.\label{fig:axial}}
\end{center}
\end{figure*}

\begin{figure}[htb]
\begin{center}
\epsfig{file=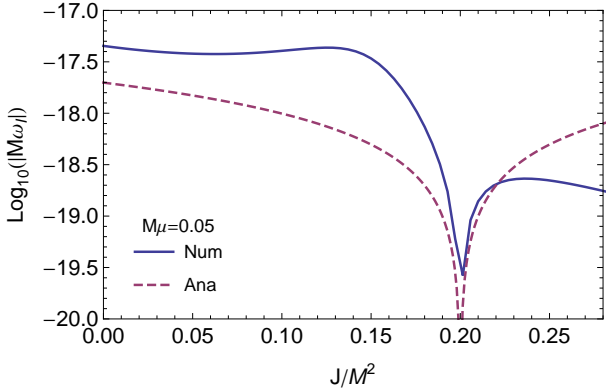,width=8cm,angle=0,clip=true}
\caption{Comparison between the numerical and analytical results for the axial mode $l=m=1$, $n=0$ as a function of the BH rotation rate $\tilde{a}$ for a mass coupling of $M\mu=0.05$. The solid line shows the numerical data and the dashed shows the analytical formula.\label{fig:ana_vs_num}}
\end{center}
\end{figure}

In the top panels of Fig.~\ref{fig:axial} we show the absolute value of the imaginary part as a function of the rotation parameter for the axial modes $l=1$, $S=1$ and $l=2$, $S=-1$. Although a second-order approximation would be needed to describe the superradiant regime in a self-consistent way~\cite{Pani:2012vp}, the first-order approximation predicts very well the onset of the instability and should give the correct order of magnitude of the instability timescale. For axial modes the instability is very weak: even in the most favorable cases the instability is almost five orders of magnitude weaker than that associated to axial Proca modes~\cite{Pani:2012bp,Pani:2012vp}. This also  makes it difficult to track numerically the axial spin-2 modes with sufficient precision. For small masses the real part of the frequency is roughly independent on the spin.  
This is supported by analytical results for the axial dipole mode, which can be evaluated analytically in the small-mass limit at first order in $\tilde{a}$ [cf. Appendix~\ref{app:ana}]. The analytical formula for the imaginary part of the fundamental mode reads
\be
M\omega_I\approx \frac{40}{19683}(\tilde{a}-2r_+\mu)(M\mu) ^{11}\,.
\ee
In Fig.~\ref{fig:ana_vs_num} we compare the analytical formula with the numerical results for the fundamental overtone and mass coupling $M\mu=0.05$. Although the imaginary part is tiny, the agreement is good in the $\mu\to0$ limit. Near the superradiant regime the agreement is only qualitative, as expected since the analytical formula is only valid for $\tilde{a}m/(M\mu)\lesssim l$.

The bottom panels of Figure~\ref{fig:axial} show the imaginary part as a function of the BH angular momentum for the polar dipole $l=1$ and the polar mode $l=2$, $S=-2$. In this case the imaginary part of the mode is larger, and these modes are easier to evaluate numerically. The instability for the mode $l=2$, $S=-2$ is roughly two orders of magnitude weaker than the strongest instability of a Proca field~\cite{Pani:2012bp}. Once more the polar dipole mode is the most interesting case as it has the largest imaginary part, corresponding to an extremely short instability timescale. This agrees with the analysis in the nonrotating case of Sec.~\ref{sec:schwar}, where we found that the behavior of this mode is different from the rest of the spectrum.

\begin{figure}[htb]
\begin{center}
\epsfig{file=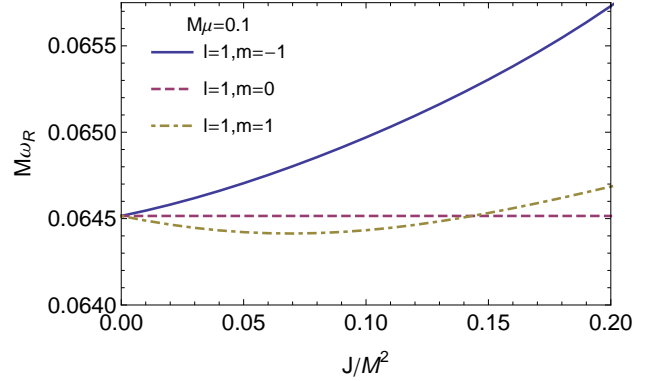,width=8.3cm,angle=0,clip=true}
\caption{Real part of the polar dipole quasibound mode as a function of the BH rotation rate $\tilde{a}$ for different values of the azimuthal number $m$ and mass coupling $\mu M=0.1$, computed at first order.\label{fig:polar_Re_dipole}}
\end{center}
\end{figure}

As shown in the bottom panels of Fig.~\ref{fig:axial}, the polar dipole mode displays a peculiar behavior in the superradiant regime, where the power-law dependence is inverted, i.e., the instability is stronger for the lowest mass coupling $M\mu$. This suggests that extrapolating the first-order results to the superradiant case is probably less accurate for this mode.
This is confirmed by the behavior of the real-part of the frequency as a function of the spin, as shown in Fig.~\ref{fig:polar_Re_dipole}. At first-order the eigenfrequencies can be expanded as
\be
\omega_R=\omega_0+\tilde{a}m\omega_1+\mathcal{O}(\tilde{a}^2)\,,
\ee
where $\omega_0$ is the eigenfrequency in the nonrotating space-time and $\omega_1$ is the first-order correction which is an even function of $m$~\cite{Pani:2012vp}. Hence at first-order we would expect that the curves for $l=m$ and $l=-m$ are symmetric when reflected around the $m=0$ curve. For the polar dipole this only happens for very small masses. Note also that, contrarily to the rest of the spectrum, the real part of the polar dipole mode acquires a nonnegligible dependence on $\tilde{a}$, even in the small $\mu$ limit. In fact the analytical results for the axial dipole suggest that the first-order approximation is only valid for $\tilde{a}m/(M\omega_R)\lesssim l$. Since in this case $M\omega_R$ is much smaller that $M\mu$, the extrapolation to the superradiant regime is less accurate in the polar dipole case. Nonetheless, using the exact results in the nonrotating case [cf. Sec.~\ref{sec:schwar}] and a linear extrapolation of the first-order corrections, we estimate the following scaling for the imaginary part of the polar dipole mode:
\be
M\omega_I\sim \gamma_{{\rm polar}}(\tilde{a}m-2r_+\omega_R)(M\mu)^{3}\,,~\label{wIpoldip}
\ee
where $\gamma_{{\rm polar}}\sim{\cal O}(1)$ and $\omega_R$ is the zeroth order real frequency given by Eq.~\eqref{polar_di_Re}. This behavior becomes less accurate deep inside the superradiant regime. Although such extrapolation is extremely rough, a similar estimate has been done in the scalar and in the Proca case and it turned out to be very accurate~\cite{Pani:2012bp}. In the scalar case a fit similar to Eq.~\eqref{wIpoldip} agrees with exact results (obtained solving the Klein-Gordon equation on an exact Kerr metric~\cite{Dolan:2007mj}) within a few percents; and, in the Proca case, it reproduces the results of exact numerical simulations (again in the quasiextremal, $\tilde{a}\sim0.99$ case) within a factor two~\cite{Witek:2012tr}. 

In the case at hand, even if Eq.~\eqref{wIpoldip} eventually turns out to be accurate only at the order-of-magnitude level, this would anyway mean that spin-2 fields can trigger the strongest superradiant instability among other bosonic perturbations.
The instability timescale is four orders of magnitude shorter than the shortest timescale for Proca unstable modes~\cite{Pani:2012bp}. A second-order analysis would be important to confirm this result, but it will also be very challenging. A most promising extension is to perform a full numerical analysis (along the lines of Ref.~\cite{Witek:2012tr}) in the case of massive spin-2 fields around highly spinning Kerr BHs.
%
%

\section{Discussion}\label{sec:conclusion}
The advent of new and powerful methods in BH perturbation theory and Numerical Relativity in the past few years 
allows one to finally tackle traditionally complex problems. Particularly important to beyond-the-Standard-Model 
physics are scenarios where ultralight bosonic degrees of freedom are present; simultaneously, massive degrees
of freedom turn out to be important outside particle physics, in particular
several extensions of general relativity encompassing massive mediators have been proposed.
Thus, the study of massive fluctuations around BHs is a timely topic.

Interesting nonlinear completions of the Fierz-Pauli theory have recently been put forward~\cite{deRham:2010ik,deRham:2010kj,Hassan:2011hr}.
While it is at this stage too early to claim a consistent theory of massive gravitons (these theories or at least certain sectors are either
pathological~\cite{2013arXiv1302.4367C,Deser:2012qx} or phenomenologically disfavored~\cite{Burrage:2012ja}), any nonlinear theory describing a massive spin-2 field --~including a massive graviton~-- will eventually reduce to Eqs.~\eqref{eqmotioncurved}--\eqref{constraint2} in the linearized regime.

Here we have explored the propagation of massive tensors in BH backgrounds as described by Eqs.~\eqref{eqmotioncurved}--\eqref{constraint2}, and shown that they lead to generic instabilities.
Schwarzschild and Kerr BHs are both unstable against linearized monopole perturbations. These are strong, small-mass instabilities
whose end-state is unknown.

Schwarzschild BHs also admit a very rich spectrum of long-lived stable states. Once rotation is turned on, these long-lived states can grow exponentially
and extract angular momentum away from the BH. Thus Kerr BHs are also unstable against a second mechanism: superradiance.
We showed that the instability is triggered when the superradiant condition is met, thus providing one further and strong piece of solid evidence that superradiant instabilities occur for any bosonic massive field. 
The polar gravitational sector is particularly interesting, as it displays the shortest instability timescale among other bosonic fields.
Our results are formally only valid in the small BH rotation limit, but previous second-order calculations for massive vector
fields suggest that a first-order analysis provides reasonably accurate results even beyond its regime of validity. The most crucial point in this regard
is the functional dependence of the instability timescale for the supposedly more unstable polar dipole mode, which we estimate to be:
\be
\tau_{{\rm tensor}}=\omega_I^{-1}\sim \frac{M(M\mu)^{-3}}{\gamma_{{\rm polar}}(\tilde{a}-2r_+\omega_R)}\,.
\ee
This timescale is four orders of magnitude shorter than the corresponding Proca field instability \cite{Pani:2012bp,Pani:2012vp}. 

It has been shown that BH superradiant instabilities together with supermassive BH spin measurements
can be used to impose stringent constraints on
the allowed mass range of massive fields \cite{Pani:2012bp,Pani:2012vp}. The observation of spinning BHs 
implies that the instability timescale is larger than typical competing spin-up effects.
For supermassive BHs a conservative estimate of these timescales is given by the Salpeter
timescale for accretion at the Eddington rate, $\tau_S\sim 4.5\times 10^7$ years.
We find that the current best bound comes from Fairall 9 \cite{Schmoll:2009gq}, for which the polar instability implies a conservative
bound $\mu\lesssim 5\times 10^{-23} {\rm eV}$.
Unlike bounds for hypothetical massive photons, which may interact strongly with matter, the previous bound should not be strongly affected by the presence of
accretion disks around BHs, as the coupling of gravitons and other spin-2 fields to matter is very feeble.

Our work requires extensions and further analysis (in particular, the understanding of the time-development of the monopole and superradiant instability
requires nonlinear simulations), and should in fact be looked at as 
the first step in a broader program of understanding gravitational-wave emission in massive theories of gravity. 
\begin{acknowledgments}
 We thank Jorge Pullin for suggesting and getting us interested in this problem, and Eugeny Babichev, Stanley Deser, Alessandro Fabbri, Nemanja Kaloper
 and Antonio Padilla for interesting correspondence.
V.C. acknowledges partial financial
support provided under the European Union's FP7 ERC Starting Grant ``The dynamics of black holes:
testing the limits of Einstein's theory'' grant agreement no.
DyBHo--256667, the NRHEP 295189 FP7-PEOPLE-2011-IRSES Grant, and FCT-Portugal through projects
PTDC/FIS/116625/2010, CERN/FP/116341/2010 and CERN/FP/123593/2011.
Research at Perimeter Institute is supported by the Government of Canada 
through Industry Canada and by the Province of Ontario through the Ministry
of Economic Development and Innovation.
R.B. acknowledges financial support from the FCT-IDPASC program through the grant SFRH/BD/52047/2012.
P.P. acknowledges financial support provided by the
European Community through the Intra-European Marie Curie contract
aStronGR-2011-298297.
\end{acknowledgments}

\appendix
\section{Linearized field equations for a spin-2 field on a Schwarzschild geometry}\label{app:decomposition}
A massive spin-2 field propagates five helicity states and one cannot impose the same gauge choices that are usually imposed in the massless case. In particular, the standard Regge-Wheeler gauge~\cite{Regge:1957td} is too restrictive for a massive spin-2 field.

In this paper, we have decomposed the spin-2 field in terms of axial and polar perturbations and expanded in a complete basis of tensor spherical harmonics. Given the expansion~\eqref{decom}, the axial and polar parts are given respectively by
\begin{widetext}
\begin{equation}\label{oddpart}
h^{{\rm axial},lm}_{\mu\nu}(\omega,r,\theta,\phi) =
 \begin{pmatrix}
  0 & 0 & h^{lm}_0(\omega,r)\csc\theta\partial_{\phi}Y_{lm}(\theta,\phi) & -h^{lm}_0(\omega,r)\sin\theta\partial_{\theta}Y_{lm}(\theta,\phi) \\
  * & 0 & h^{lm}_1(\omega,r)\csc\theta\partial_{\phi}Y_{lm}(\theta,\phi) & -h^{lm}_1(\omega,r)\sin\theta\partial_{\theta}Y_{lm}(\theta,\phi) \\
  *  & *  & -h^{lm}_2(\omega,r)\frac{X_{lm}(\theta,\phi)}{\sin\theta} & h^{lm}_2(\omega,r)\sin\theta W_{lm}(\theta,\phi)  \\
  * & * & * & h^{lm}_2(\omega,r)\sin\theta X_{lm}(\theta,\phi)
 \end{pmatrix}\,,
\end{equation}
%
%
\begin{align}\label{evenpart}
h^{{\rm polar},lm}_{\mu\nu}(\omega,r,\theta,\phi)=
\begin{pmatrix}
f(r)H_0^{lm}(\omega,r)Y_{lm} & H_1^{lm}(\omega,r)Y_{lm} & \eta^{lm}_0(\omega,r)\partial_{\theta}Y_{lm}& \eta^{lm}_0(\omega,r)\partial_{\phi}Y_{lm}\\
  * & f(r)^{-1} H_2^{lm}(\omega,r)Y_{lm} & \eta^{lm}_1(\omega,r)\partial_{\theta}Y_{lm}& \eta^{lm}_1(\omega,r)\partial_{\phi}Y_{lm}\\
  *  & *  & \begin{array}{c}r^2\left[K^{lm}(\omega,r)Y_{lm}\right.\\\left. +G^{lm}(\omega,r)W_{lm}\right]\end{array} & r^2  G^{lm}(\omega,r)X_{lm}  \\
  * & * & * & \begin{array}{c}r^2\sin^2\theta\left[K^{lm}(\omega,r)Y_{lm}\right.\\\left.-G^{lm}(\omega,r)W_{lm}\right]\end{array}
\end{pmatrix}\,,
\end{align}
\end{widetext}
where $f(r)=1-2M/r$, asterisks represent symmetric components, $Y_{lm}\equiv Y_{lm}(\theta,\phi)$ are the scalar spherical harmonics and

\be
X_{lm}(\theta,\phi)=2\partial_{\phi}\left[\partial_{\theta}Y_{lm}-\cot\theta Y_{lm}\right]\,,
\ee
\be
W_{lm}(\theta,\phi)=\partial^2_{\theta}Y_{lm}-\cot\theta\partial_{\theta}Y_{lm}-\csc^2\theta\partial^2_{\phi}Y_{lm}\,.
\ee
%

\subsection{Axial equations} 
The field equations for the axial sector are obtained
by using the decomposition \eqref{oddpart} in Eq.~\eqref{eqmotioncurved}. Substituting into the linearized field equations, we obtain:
\beq
&&f^2 h''_0+\left[\omega^2-f\left(\mu^2+\frac{\Lambda}{r^2}-\frac{4M}{r^3}\right)\right]h_0\nn\\
&&-\frac{2Mi\omega f}{r^2} h_1=0\,,\label{odd1}\\
&&f^2 h''_1+\frac{4 M f}{r^2}h'_1+\left[\omega^2-f\left(\mu^2+\frac{\Lambda+4}{r^2}-\frac{8M}{r^3}\right)\right]h_1\nn\\
&&-\frac{2Mi\omega}{r(r-2M)}h_0+\frac{2(2-\Lambda)f}{r^3}h_2=0\,,\label{odd2}\\
&&f^2 h''_2-\frac{2 f(r-3M)}{r^2}h'_2-\frac{2f^2}{r}h_1\nn\\
&&+\left[\omega^2-f\left(\mu^2+\frac{\Lambda-4}{r^2}+\frac{8M}{r^3}\right)\right]h_2=0\,,\label{odd3}
\eeq
where $\Lambda=l(l+1)$ and $f\equiv f(r)$. Equations \eqref{odd1} and \eqref{odd2} correspond to the $(t\theta)$ and the $(r\theta)$
component of the field equations respectively, and \eqref{odd3} corresponds to the $(\theta\theta)$ component.
The transverse constraint \eqref{constraint1} leads to the radial equation
\begin{equation}
\label{harmonicodd}
f h'_1-\frac{2 (M-r)}{r^2}h_1+\frac{i\omega}{f}h_0+\frac{\Lambda-2}{r^2}h_2=0\,,
\end{equation}
which can be obtained either from the $\theta$ or the $\phi$ component.
For the axial terms the trace~\eqref{constraint2} vanishes identically,
\begin{equation}
\label{traceodd}
h^{\rm axial}=0\,.
\end{equation}
Using the constraint \eqref{harmonicodd} we can reduce the system to a pair of coupled differential equations. Eliminating $h_0$, we finally obtain the system~\eqref{oddf1}-\eqref{oddf2}.

\subsection{Polar equations} 
Using the decomposition \eqref{evenpart} in Eq.~\eqref{eqmotioncurved} and substituting into the linearized field equations, we obtain:
\begin{widetext}
\begin{flalign}\label{even1}
f^2 H''_0&+\frac{2 f(r-M)}{r^2}H'_0+\left[\omega^2-\frac{2M^2}{r^4}-f\left(\mu^2+\frac{\Lambda}{r^2}\right)\right]H_0-\frac{4iM\omega}{r^2}H_1-\frac{2M(2r-3M)}{r^4}H_2+\frac{4M f}{r^3}K=0\,,&
\end{flalign}
\begin{flalign}
\label{even2}
f^2 H''_1&+\frac{2 f(r-M)}{r^2}H'_1+\left[\omega^2-\frac{4M^2}{r^4}-f\left(\mu^2+\frac{\Lambda+2}{r^2}\right)\right]H_1-\frac{2iM\omega}{r^2}(H_0+H_2)+\frac{2\Lambda f}{r^3}\eta_0=0\,,&\end{flalign}
\begin{flalign}
\label{even3}
f^2 H''_2&+\frac{2 f(r-M)}{r^2}H'_2+\left[\omega^2-\frac{2M^2}{r^4}-f\left(\mu^2+\frac{\Lambda+4}{r^2}-\frac{8M}{r^3}\right)\right]H_2-\frac{2M(2r-3M)}{r^4}H_0-\frac{4iM\omega}{r^2}H_1&\nn\\
&+\frac{4(r-3M)f}{r^3}K+\frac{4\Lambda f^2}{r^3}\eta_1=0\,,&
\end{flalign}
\begin{flalign}
\label{even4}
f^2 \eta''_0&+\left[\omega^2-f\left(\mu^2+\frac{\Lambda}{r^2}-\frac{4M}{r^3}\right)\right]\eta_0-\frac{2M i\omega f}{r^2}\eta_1+\frac{2 f^2}{r}H_1=0\,,&
\end{flalign}
\begin{flalign}
\label{even5}
f^2 \eta''_1&+\frac{4M f}{r^2}\eta'_1+\left[\omega^2-f\left(\mu^2+\frac{\Lambda+4}{r^2}-\frac{8M}{r^3}\right)\right]\eta_1-\frac{2M i\omega}{r(r-2M)}\eta_0+\frac{2 f}{r}\left[H_2-K+(\Lambda-2)G\right]=0\,,&
\end{flalign}
\begin{flalign}
\label{even6}
f^2 G''&+\frac{2(r-M) f}{r^2}G'+\left[\omega^2-f\left(\mu^2+\frac{\Lambda-2}{r^2}\right)\right]G+\frac{2f^2}{r^3}\eta_1=0\,,&
\end{flalign}
\begin{flalign}
\label{even7}
f^2 K''&+\frac{2(r-M) f}{r^2}K'+\left[\omega^2-f\left(\mu^2+\frac{\Lambda+2}{r^2}-\frac{8M}{r^3}\right)\right]K+\frac{2M f}{r^3}H_0+\frac{2(r-3M) f}{r^3}H_2-\frac{2\Lambda f^2}{r^3}\eta_1=0\,,&
\end{flalign}
\end{widetext}
Equations \eqref{even1}-\eqref{even5} correspond to the $(tt),\,(tr),\,(rr),\,(t\theta)$ and $(r\theta)$ components of the field equations, respectively. From the $(\theta\phi)$ component we get Eq.~\eqref{even6}, which combined with the $(\theta\theta)$
component yields Eq.~\eqref{even7}.

The transverse constraint \eqref{constraint1} leads to the following radial equations
\be
\label{harmoniceven1}
f H'_1-\frac{2 (M-r)}{r^2}H_1+i\omega H_0-\frac{\Lambda}{r^2}\eta_0=0\,,
\ee
\be
\label{harmoniceven2}
f H'_2+\frac{2r-3M}{r^2}H_2+i\omega H_1+\frac{M}{r^2}H_0-\frac{2f}{r}K-\frac{f\Lambda}{r^2}\eta_1=0\,,
\ee
\be
\label{harmoniceven3}
f \eta'_1-\frac{2(M-r)}{r^2}\eta_1+\frac{i\omega}{f} \eta_0+K-(\Lambda-2)G=0\,,
\ee
for the $t,r$ and $\theta$ component of the constraint, respectively.
Finally, in the polar case the traceless constraint~\eqref{constraint2} yields
\be
H_0=H_2+2 K\,.\label{traceeven}
\ee
%
Unlike the axial sector, the polar equations are not so straightforward to further reduce. For $l\geq 2$ one could use the constraint equations to eliminate $\eta_0$, $\eta_1$, $H_0$ and $G$ and obtain three second-order equations for $K$, $H_1$ and $H_2$. However, this choice is not particularly  useful, because the system does not directly contain the monopole and dipole cases ($l=0,1$). For this reason we chose to work with $K$, $\eta_1$ and $G$ as dynamical variables instead.

After some tedious algebra, we obtain that the polar sector is fully described by Eqs.~\eqref{polar_eq1}--\eqref{polar_eq3} in the main text.

\section{Eigenvalue problem: Quasinormal modes and quasibound states}\label{app:modes}
This appendix details the numerical computation of BH eigenfrequencies for massive perturbations.
To have a well-defined problem we need to define boundary conditions, and these
determine an eigenvalue problem for the frequency $\omega$, which can be solved using 
several different tools~\cite{Berti:2009kk,PaniNRHEP2}. 
At the horizon we must impose regular boundary conditions, which correspond to purely ingoing waves
\be
\label{BC_hor}
\Phi_j(r)\sim e^{-i\omega r_*}\,,\qquad j=1,2,\ldots,10\,,
\ee
as $r_*\to -\infty$, where $\Phi_j(r)$ is any of the radial functions. 
On the other hand, the asymptotic behavior of the solution at infinity is given by
\be
\label{BC_inf}
\Phi_j(r)\sim B_j e^{-i k_{\infty} r}r^{-\frac{M(\mu^2-2\omega^2)}{k_{\infty}}}+C_j e^{i k_{\infty} r}r^{\frac{M(\mu^2-2\omega^2)}{k_{\infty}}}\,,
\ee
where $k_{\infty}=\sqrt{\mu^2-\omega^2}$, such that Re$(k_{\infty})>0$. For massive fields we have to consider two kinds of modes: (i) the quasinormal modes (QNM), which corresponds to purely outgoing waves at infinity, i.e., they are defined by $B_j=0$; (ii) quasibound states, defined by $C_j=0$ and correspond to modes spatially localized within the vicinity of the BH and that decay exponentially at spatial infinity.

\subsubsection{Continued-fraction method}
The use of the continued fraction method requires a suitable \emph{ansatz}, which we take to be
\begin{equation}
\label{ansatz}
\Phi_j(\omega,r)=f(r)^{-2iM\omega}r^{\nu}e^{-qr}\sum_n{a^{(j)}_n}f(r)^n\,,
\end{equation} 
where $\nu$ and $q$ are defined as below Eq.~\eqref{ansatz2}.
\subsubsection{Axial dipole}
Inserting \eqref{ansatz} into \eqref{oddl1} leads to a three-term recurrence relation of the form 
\beq
\alpha_0 a_1+\beta_0 a_0&=&0\,,\nonumber\\
\alpha_n a_{n+1}+\beta_n a_n+\gamma_n a_{n-1}&=&0\,,\qquad n>0\,,
\eeq
where,
\begin{align}
\alpha_n &= (n+1)(n+1-4i\omega)\,,\\
\beta_n &= -2 \left(n^2+n-1\right)+\frac{\omega ^2 (2 n-4 i \omega +1)}{q}\nn\\
&-3 q (2 n-4 i \omega +1)+4 i (2 n+1) \omega -4 q^2+12 \omega ^2\,,\\
\gamma_n &= q^{-2}\left(nq+q^2-3q-2 i q \omega -\omega ^2\right)\nn\\
&\times\left(nq+q^2+3q-2 i q \omega -\omega ^2\right)\,.
\end{align}
The QNM or quasibound-state frequencies can be obtained solving numerically the continued fraction equation
\be
\beta_0-\frac{\alpha_0 \gamma_1}{\beta_1-\frac{\alpha_1 \gamma_2}{\beta_2-\frac{\alpha_2 \gamma_3}{\beta_3-\ldots}}}=0\,.
\ee
This method has been extensively used and described in detail elsewhere~\cite{Leaver:1985ax,Berti:2009kk,Pani:2012bp}, some routines are freely available \cite{webpage} so we will not discuss it any further. 
\subsubsection{Axial modes: $l\geq 2$}
For $l\geq 2$ the axial modes satisfy a pair of coupled differential equations, Eqs.~\eqref{oddf1} and \eqref{oddf2}. Inserting \eqref{ansatz} into these equations leads to a three-term matrix-valued recurrence relation,
\begin{eqnarray}
\label{recrelation}
\bm{\alpha}_0 \mathbf{U}_1+\bm{\beta}_0 \mathbf{U}_0&=&0\,,\nonumber\\
\bm{\alpha}_n \mathbf{U}_{n+1}+\bm{\beta}_n \mathbf{U}_n+\bm{\gamma}_n \mathbf{U}_{n-1}&=&0\,,\qquad n>0\,,
\end{eqnarray}
The quantity $\mathbf{U}_n=\left(a_n^{(1)},a_n^{(2)}\right)$ is a two-dimensional vectorial coefficient and $\bm{\alpha}_n$, $\bm{\beta}_n$, $\bm{\gamma}_n$ are $2\times 2$ matrices whose form reads,
\beq
\bm{\alpha}_n &=&
\begin{pmatrix}
\alpha_n & 0 \\
   0     & \alpha_n 
\end{pmatrix}\,, \qquad
\bm{\beta}_n =
\begin{pmatrix}
\beta_n & \Lambda-2 \\
   -2   & \beta_n-3 
 \end{pmatrix}\,, \nonumber\\
\bm{\gamma}_n &=&
\begin{pmatrix}
  \gamma_n & 6-3\Lambda \\
   0 & \gamma_n+9 
\end{pmatrix}\,, \nonumber
\eeq
with
\begin{align}
\alpha_n &= (n+1)(n+1-4i\omega)\,,\\
\beta_n &= 2-\Lambda -2 \left(n^2+n-1\right)+\frac{\omega ^2 (2 n-4 i \omega +1)}{q}\nn\\
&-3 q (2 n-4 i \omega +1)+4 i (2 n+1) \omega -4 q^2+12 \omega ^2\,,\\
\gamma_n &= q^{-2}\left[q^2 \left(n^2-4 i n \omega -6 \omega ^2-9\right)+2 q^3 (n-2 i \omega )\right.\nn\\
&\left.-2 q \omega ^2 (n-2 i \omega )+q^4+\omega ^4\right]\,.
\end{align}
The matrix-valued three-term recurrence relation can be solved using matrix-valued continued fractions~\cite{Rosa:2011my,Pani:2012bp}. The QNM or quasibound frequencies are roots of the equation $\mathbf{M}\mathbf{U}_0=0$, where
\be
\mathbf{M}\equiv \bm{\beta}_0+\bm{\alpha}_0 \mathbf{R}^{\dagger}_0\,,
\ee
with $\mathbf{U}_{n+1}=\mathbf{R}^{\dagger}_n \mathbf{U}_n$ and
\be
\mathbf{R}^{\dagger}_n=-\left(\bm{\beta}_{n+1}+\bm{\alpha}_{n+1}\mathbf{R}^{\dagger}_{n+1}\right)^{-1} \bm{\gamma}_{n+1}\,.
\ee
For nontrivial solutions we then solve numerically
\be
\det |\mathbf{M}|=0\,.
\ee
%
\subsubsection{Direct integration for quasibound states}
To compute the spectrum of quasibound states a direct integration approach is often possible, since the solutions asymptotically vanish at spatial infinity, and desirable because it converges faster. We start with a series expansion close to the horizon of the form
\be
\label{bc_di}
\Phi_j(\omega,r)=e^{-i\omega r_*}\sum_n{b^{(j)}_n}(r-r_H)^n\,,
\ee
where the coefficients $b^{(j)}_n$ for $n\geq 1$ can be found in terms of $b^{(j)}_0$ by solving the near-horizon equations order by order.
We then integrate outward up to infinity where the condition $C_j=0$ in Eq.~\eqref{BC_inf} is imposed. This allow us to obtain the frequency spectrum using a shooting method. This method can be extended to solve systems of coupled equations~\cite{Rosa:2011my,Pani:2012bp}. Consider a system of $N$ coupled equations. Imposing the ingoing wave boundary condition at the horizon~\eqref{bc_di} we may obtain a family of solutions at infinity characterized by $N$ parameters, corresponding to the $N$-dimensional vector of the coefficients $\bm{b_0}=\{b^{(j)}_0\}$, with $j=1,\ldots,N$. Note that all the solutions of the system of coupled equations must have the form~\eqref{bc_di} near the horizon. We may then compute the bound-state spectrum by choosing a suitable orthogonal basis for the space of initial coefficients $b^{(j)}_0$. To do so we perform $N$ integrations from the horizon to infinity and construct the $N\times N$ matrix 
\be
\label{detS}
\bm{S_m} (\omega)=\lim_{r\to \infty}
 \begin{pmatrix}
  \Phi_{(1)}^{(1)} & \Phi_{(1)}^{(2)} & \ldots & \Phi_{(1)}^{(N)} \\
  \Phi_{(2)}^{(1)} & \Phi_{(2)}^{(2)} & \ldots & \ldots \\
  \ldots  & \ldots  & \ldots & \ldots  \\
  \Phi_{(N)}^{(1)} & \ldots & \ldots & \Phi_{(N)}^{(N)}
 \end{pmatrix}\,,
\ee
where the superscripts denote a particular vector of the chosen basis, for example, $\Phi_j^{(1)}$ corresponds to $\bm{b_0}=\{1,0,\ldots,0\}$, $\Phi_j^{(2)}$ corresponds to $\bm{b_0}=\{0,1,\ldots,0\}$, and $\Phi_j^{(N)}$ corresponds to $\bm{b_0}=\{0,0,\ldots,1\}$. The bound-state frequency $\omega_0=\omega_R+i\omega_I$ will then correspond to the solutions of
\be
\det|\bm{S_m}(\omega_0)|=0\,,
\ee
which in practice corresponds to minimizing $\det\bm{S_m}$ in the complex plane at arbitrarily large distances.

\section{Linearized field equations for a spin-2 field on a slowly rotating Kerr BH}\label{app:kerr}
We will follow Kojima~\cite{Kojima:1992ie} to write the fields equations for a spin-2 field in a slowly rotating BH. Since this background is still ``almost'' spherically symmetric we can use the decomposition \eqref{decom} and insert it in the linearized field equations. We can then separate the equations in three different groups.

From the $(tt)$, $(tr)$, $(rr)$, the sum of $(\theta\theta)$ and $(\phi\phi)$ components of Eq.~\eqref{eqmotioncurved}, the $t$ and $r$ components of the transverse condition~\eqref{constraint1}, and the traceless condition~\eqref{constraint2}, we have 
\begin{align}
\label{eqscalar}
&\left(A_{lm}^{(I)}+\tilde{A}_{lm}^{(I)}\cos\theta\right)Y^{lm}+B_{lm}^{(I)}\sin\theta \partial_{\theta} Y^{lm}\nn\\
&+C_{lm}^{(I)}\partial_{\phi}Y^{lm}=0\quad (I=0,\,\dots\,,6)\,,
\end{align}
where a sum over ($l,m$) is implicit, the functions $A_{lm}^{(I)}$ and $C_{lm}^{(I)}$ are some linear combinations of the polar functions $H_0$, $H_1$,$H_2$, $\eta_0$, $\eta_1$, $K$ and $G$. On the other hand $\tilde{A}_{lm}^{(I)}$ and $B_{lm}^{(I)}$ are some linear combinations of the axial functions $h_0$, $h_1$, $h_2$. 

From the $(t\theta)$, $(t\phi)$, $(r\theta)$, $(r\phi)$ components of Eq.~\eqref{eqmotioncurved}, and the $\theta$, $\phi$ components of Eq.~\eqref{constraint1}, we have 
\begin{align}
\label{eqvector1}
&\left(\alpha_{lm}^{(J)}+\tilde{\alpha}_{lm}^{(J)}\cos\theta\right)\partial_{\theta}Y^{lm}\nn\\
&-\left(\beta_{lm}^{(J)}+\tilde{\beta}_{lm}^{(J)}\cos\theta\right)\left(\partial_{\phi}Y^{lm}/\sin\theta\right)+\eta_{lm}^{(J)}(\sin\theta Y^{lm})\nn\\
&+\xi_{lm}^{(J)}X^{lm}+\chi_{lm}^{(J)}(\sin\theta W^{lm})=0 \quad (J=0,1,2)\,,
\end{align}
and
\begin{align}
\label{eqvector2}
&\left(\beta_{lm}^{(J)}+\tilde{\beta}_{lm}^{(J)}\cos\theta\right)\partial_{\theta}Y^{lm}\nn\\
&+\left(\alpha_{lm}^{(J)}+\tilde{\alpha}_{lm}^{(J)}\cos\theta\right)\left(\partial_{\phi}Y^{lm}/\sin\theta\right)+\zeta_{lm}^{(J)}(\sin\theta Y^{lm})\nn\\
&+\chi_{lm}^{(J)}X^{lm}-\xi_{lm}^{(J)}(\sin\theta W^{lm})=0\quad (J=0,1,2)\,,
\end{align}
where the functions $\alpha_{lm}^{(J)}$, $\tilde{\beta}_{lm}^{(J)}$, $\zeta_{lm}^{(J)}$ and $\xi_{lm}^{(J)}$ are some linear combination of the polar functions, while  $\beta_{lm}^{(J)}$, $\tilde{\alpha}_{lm}^{(J)}$, $\eta_{lm}^{(J)}$ and $\chi_{lm}^{(J)}$ belong to the axial sector.

From the $(\theta\phi)$ and the subtraction of $(\theta\theta)$ and $(\phi\phi)$ components of~\eqref{eqmotioncurved}, we have
\begin{align}
\label{eqtensor1}
&f_{lm}\partial_{\theta}Y^{lm}+g_{lm}\left(\partial_{\phi}Y^{lm}/\sin\theta\right)\nn\\
&+\left(s_{lm}+\hat{s}_{lm}\partial_{\phi}\right)\left(X^{lm}/\sin^2 \theta \right)\nn\\
&+\left(t_{lm}+\hat{t}_{lm}\partial_{\phi}\right)\left(W^{lm}/\sin\theta\right)=0\,,
\end{align}
and
\begin{align}
\label{eqtensor2}
&g_{lm}\partial_{\theta}Y^{lm}-f_{lm}\left(\partial_{\phi}Y^{lm}/\sin\theta\right)\nn\\
&-\left(t_{lm}+\hat{t}_{lm}\partial_{\phi}\right)\left(X^{lm}/\sin^2 \theta\right)\nn\\
&+\left(s_{lm}+\hat{s}_{lm}\partial_{\phi}\right)\left(W^{lm}/\sin\theta\right)=0\,,
\end{align}
where $f_{lm}$, $s_{lm}$ and $\hat{s}_{lm}$ are some linear combinations of polar functions and $g_{lm}$, $t_{lm}$ and $\hat{t}_{lm}$ from the axial functions.

It is easy to see that at zeroth-order in the rotation the perturbation equations reduce to
\be
A_{lm}^{(I)}=\alpha_{lm}^{(J)}=s_{lm}=0\,,\quad (I=0,\,\dots\,,6,\,J=0,1,2)\,,
\ee
for the polar sector and to
\be
\beta_{lm}^{(J)}=t_{lm}=0\,,\quad (\,J=0,1,2)\,,
\ee
for the axial sector, respectively. These equations correspond to the ones obtained for the Schwarzschild case.

To separate the angular variables we use the identities
\begin{align}
\cos\theta Y^{lm}&= Q_{l+1\,m} Y^{l+1\,m}+Q_{lm} Y^{l-1\,m}\,,\\
\sin\theta\partial_\theta Y^{lm}&= Q_{l+1\,m}\,l\, Y^{l+1\,m}-Q_{lm}(l+1) Y^{l-1\,m}\,,
\end{align}
with
\be
Q_{lm}=\sqrt{\frac{l^2-m^2}{4l^2-1}}\,,
\ee
and the orthogonality properties of scalar, vector and tensor harmonics. 
The separation of the angular dependence of Einstein's equations for a
slowly-rotating star was performed in Ref.~\cite{Kojima:1992ie}. Since the above equations are formally the same as those considered in Ref.~\cite{Kojima:1992ie}, they can be separated in exactly the same way [see Ref.~\cite{PaniNRHEP2} for a review].
Below we omit the index $m$, because in an axisymmetric background it is possible to decouple the perturbation equations so that all quantities have the same value of $m$. 

From Eq.~\eqref{eqscalar} we have~\cite{Kojima:1992ie,PaniNRHEP2}
\begin{align}
\label{descalar}
&A_{l}^{(I)}+i m C_{l}^{(I)}+Q_{l}\left(\tilde{A}_{l-1}^{(I)}+(l-1)B_{l-1}^{(I)}\right)\nn\\
&+Q_{l+1}\left(\tilde{A}_{l+1}^{(I)}-(l+2)B_{l+1}^{(I)}\right)=0\,.
\end{align}

Equations~\eqref{eqvector1} and \eqref{eqvector2} give 
\begin{align}
\label{devector1}
&\Lambda\alpha_{l}^{(J)}+i m\left[(l-1)(l+2)\xi_{l}^{(J)}-\tilde{\beta}_{l}^{(J)}-\zeta_{l}^{(J)}\right]\nn\\
&+Q_l (l+1)\left[(l-2)(l-1)\chi_{l-1}^{(J)}+(l-1)\tilde{\alpha}_{l-1}^{(J)}-\eta_{l-1}^{(J)}\right]\nn\\
&-Q_{l+1}\,l\left[(l+2)(l+3)\chi_{l+1}^{(J)}-(l+2)\tilde{\alpha}_{l+1}^{(J)}-\eta_{l+1}^{(J)}\right]=0\,,
\end{align}
and
\begin{align}
\label{devector2}
&\Lambda\beta_{l}^{(J)}+i m\left[(l-1)(l+2)\chi_{l}^{(J)}+\tilde{\alpha}_{l}^{(J)}+\eta_{l}^{(J)}\right]\nn\\
&-Q_l (l+1)\left[(l-2)(l-1)\xi_{l-1}^{(J)}-(l-1)\tilde{\beta}_{l-1}^{(J)}+\zeta_{l-1}^{(J)}\right]\nn\\
&+Q_{l+1}\,l\left[(l+2)(l+3)\xi_{l+1}^{(J)}+(l+2)\tilde{\beta}_{l+1}^{(J)}+\zeta_{l+1}^{(J)}\right]=0\,.
\end{align}
Finally, Eqs.~\eqref{eqtensor1} and \eqref{eqtensor2} yield 
\be
\label{detensor1}
\Lambda \left(s_{l}+i m\hat{s}_{l}\right)-i m f_{l}-Q_l (l+1) g_{l-1}+Q_{l+1}l\,g_{l+1}=0\,,
\ee
\be
\label{detensor2}
\Lambda \left(t_{l}+i m\hat{t}_{l}\right)+i m g_{l}-Q_l (l+1) f_{l-1}+Q_{l+1}l\,f_{l+1}=0\,.
\ee

Because the background is nonspherically symmetric, the radial equations above display mixing between perturbations with opposite parity and different harmonic index. To first order, perturbations with given parity and harmonic index $l$ are coupled to perturbations with opposite parity and indices $l\pm1$. However, as discussed in Ref.~\cite{Pani:2012vp}, these couplings do not contribute to the eigenvalue spectrum to first order in $\tilde{a}$. Finally, neglecting the coupling to the opposite parity with harmonic indices $l\pm 1$, we use Eqs.~\eqref{devector2} and~\eqref{detensor2} to deduce the axial equations~\eqref{axial1} and~\eqref{axial2} in the main text, while the polar equations are obtained from Eqs.~\eqref{descalar},~\eqref{devector1} and~\eqref{detensor1}.

\section{Analytical results for the axial dipole}\label{app:ana}
In this appendix we generalize Detweiler's analytical calculations~\cite{Detweiler:1980uk} for the unstable scalar modes of a Kerr BH in the small-mass limit to the case of the massive spin-2 axial dipole, to first-order in the rotation. 

Defining $R(r)=Q/r$ the axial dipole equation~\eqref{axialdi_kerr} can be rewritten as
\begin{align}
&r^2 f\frac{d}{dr}\left(r^2 f\frac{dR}{dr}\right)+\Big[r^4\omega^2-4\tilde{a}m M^2 r\omega-r^2 f \Big( j(j+1)\nn\\
&\left.\left.+\mu^2r^2-\frac{2M s'^2}{r}-\tilde{a}m M^2\frac{12 (4 r-9 M)}{r^4 \omega}\right)\right]R=0\,,
\end{align}
where we have defined $j=l+S=2$ and $s'=3$. From now on we consider $j$ and $s'$ to be generic integers and we replace their specific values only in the final result~\eqref{finalana} below. The latter is valid for any $j$ and $s'$ provided $j<s'$. To use the method of matching asymptotics we start by writing this equation in terms of the dimensionless variable $z=(r-r_+)/r_+$, 
\begin{align}
\label{di_ana}
&Z\frac{d}{dz}\left(Z\frac{dR}{dz}\right)+\Big[4M^2\omega^2(1+z)^4-2\tilde{a}m M \omega(1+z)\nn\\
&-j(j+1)Z-4M^2\mu^2 z(1+z)^3+s'^2 z\nn\\
&\left.-\tilde{a}m\frac{3z(1-8z)}{4M \omega(1+z)^3}\right]R=0\,,
\end{align}
where $Z=z(z+1)$.

We first expand the equation above for $z\gg 1$. For this we define the variable $x=4Mk_{\infty}z$ and get the equation   
\be
\label{di_zgg1}
\frac{d^2}{dx^2}\left(x R\right)+\left[-\frac{1}{4}+\frac{\nu}{x}-\frac{j(j+1)}{x^2}\right]xR=0\,,
\ee
where we have defined , $k^2_{\infty}=\mu^2-\omega^2$, $\nu=M\mu^2/k_{\infty}$ and have considered $\omega\sim \mu$.
For quasibound states the solution of this equation with the correct boundary condition at infinity is given by
\be
R_{\infty}(x)\approx C_1 e^{-x/2}x^j U(1+j-\nu,2j+2,x)\,,
\ee
where $C_1$ is a constant and $U(p,q,x)$ is one of the confluent hypergeometric functions~\cite{handmath}. For $z\ll 1$, at leading order, the behavior of the solution reads
\begin{align}
&R_{\infty}(r)\approx C_1 \left[(2k_{\infty}r)^j\frac{\Gamma[-1-2j]}{\Gamma[-j-\nu]}\right.\nn\\
&\left.+(2k_{\infty}r)^{-j-1}\frac{\Gamma[1+2j]}{\Gamma[1+j-\nu]}\right]\,.
\end{align} 
Equation~\eqref{di_ana} can also be solved in the region where $r\ll$ max$(j/\omega,j/\mu)$. In this limit,
\be
Z\frac{d}{dz}\left(Z \frac{dR}{dz}\right)+\left[P^2-j(j+1)Z+\bar{s}^2 z\right]R=0\,,
\ee
where we have defined  $\epsilon=2M\mu$, $\bar{s}^2=s'^2-\frac{3\tilde{a}m}{2\epsilon}$, $P=-2Mk_H=-2M(\omega-m\Omega_H)$ and neglect $\mathcal{O}(\tilde{a}^2)$ terms in $P^2$. Note that in order to solve the equation analytically, we neglect terms $\mathcal{O}(\frac{\tilde{a}z^2}{\epsilon})$, so the approximation is valid only if $\tilde{a}\ll j\, M\mu$. 

The solution of the equation above is given in terms of hypergeometric functions. Imposing ingoing waves at the horizon we get that the general solution is given by
\begin{align}
&R_{H}(r)=C_2 e^{-2P\pi}(-1)^{2iP}z^{iP}(1+z)^{\sigma}\nn\\
&_2F_1(-j+i P+\sigma,1+j+i P+\sigma,1+2i P,-z)\,,
\end{align}
where $_2F_1(a,b,c,z)$ is the hypergeometric function~\cite{handmath} and $\sigma=\sqrt{\bar{s}^2-P^2}$. Using the asymptotic properties of the hypergeometric function~\cite{handmath} we can derive the large-distance limit $z\gg 1$ of this solution
\begin{align}
&R_{H}(r)\approx C_2 \Gamma[1+2iP]\nn\\
&\times\left[\frac{(2M)^{1+j}\Gamma[-1-2j]}{\Gamma[-j+i P-\sigma]\Gamma[-j+i P+\sigma]}r^{-j-1}\right.\nn\\
&\left.+\frac{(2M)^{-j}\Gamma[1+2j]}{\Gamma[1+j+i P-\sigma]\Gamma[1+j+i P+\sigma]}r^{j}\right]\,.
\end{align} 
The near- and far-region solutions have an overlapping region when $M\omega\ll j$ and $M\mu\ll j$ and one can find a matching condition equating the coefficients of $r^j$ and $r^{-j-1}$:
\begin{align}
&\frac{\Gamma [2 j+1] \Gamma [-j-\nu]}{\Gamma [-2 j-1] \Gamma [j-\nu
   +1]}=(4k_{\infty} M)^{2 j+1}\nn\\
	\times&\frac{\Gamma [-2 j-1]\Gamma \left[j+i
   P-\sigma+1\right] \Gamma \left[j+i
   P+\sigma+1\right]}{\Gamma [2 j+1] \Gamma \left[-j+i
   P-\sigma\right] \Gamma \left[-j+i P+\sigma\right]}\,.
\end{align}
At leading order for $M k_{\infty}$ the right hand side vanishes. In the left hand side this corresponds to the poles of $\Gamma[j+1-\nu]$, which are given by $\nu^{(0)}=j+1+n$ for a non-negative integer $n$, yielding the expected hydrogen-like quasibound states. We obtain, to lowest order in $M\mu$,
\be
\label{ana_real}
k_{\infty}^2=\mu^2-\omega_R^2\approx\mu^2\left(\frac{M\mu}{j+n+1}\right)^2\,.
\ee
In order to get the imaginary part of the spectrum, we expand around this value to get the next-to-leading order correction. Writing $\nu\equiv \nu^{(0)}+\delta\nu$ and assuming $\delta\nu\ll 1$ we get (for details see e.g.~\cite{Furuhashi:2004jk})
\begin{align}	
\label{ana_kerr}
&\delta\nu\approx -\frac{(4 k_{\infty} M)^{2 j+1}\Gamma [-2 j-1] \Gamma [2 j+n+2]}{\Gamma [1+2 j]^2\Gamma [2 j+2]
   \Gamma [n+1] } \times\nn\\
&	\frac{\Gamma [j+iP-\sigma +1] \Gamma [j+i P+\sigma +1]}{\Gamma [-j+i P-\sigma ] \Gamma [-j+i P+\sigma ]}\,.
\end{align}
Since there is a pole in one of the $\Gamma$-functions we take to lowest order in $P$ and $\tilde{a}/\epsilon$, $\Gamma[-j+i P-\sigma]\approx \Gamma[-j-s']$.
%
%
We then get in this limit
\begin{align}
\delta\nu&\approx (-1)^{j-s'}\frac{(4 k_{\infty} M)^{2 j+1} \Gamma [2 j+n+2]\Gamma[j+s'+1]^2}{2\Gamma [1+2 j]^2\Gamma [2 j+2]^2
   \Gamma [n+1] \Gamma [-j+s']}\nn\\
	&\times\Gamma [j+iP-\sigma +1] \,,
\end{align}
where the factor $2$ in the denominator comes from a specific limit of the $\Gamma$ functions and it is related to the fact that $l(l+1)$ is not the exact angular eigenvalue in a rotating background. In the nonrotating limit, the result above must be multiplied by a factor $2$. [see the discussion of Appendix C2 in Ref.~\cite{Pani:2012vp} for details].
The imaginary part of the bound-mode frequency reads
\be
i\omega_I=\frac{\delta\nu}{M}\left(\frac{M\mu}{j+n+1}\right)^3\,.
\ee
To understand how this scales with $M\mu$ in the small-mass limit, we note that for $\tilde{a}\ll M\mu$  and at first-order in $P$ we have  $\Gamma [j+iP-\sigma +1]\sim -i P/{P^2}\sim -\frac{iP}{4M^2\mu^2}$. Finally we get 
\begin{align}
&M\omega_I\approx (-1)^{j+1-s'}(\tilde{a}m-2r_+\mu)(M\mu)^{4j+3}\times\nn\\
&\frac{4^{2j-1} \Gamma [2 j+n+2]\Gamma[j+s'+1]^2}{(j+1+n)^{2j+4}\Gamma [1+2 j]^2\Gamma [2+2 j]^2
   \Gamma [n+1] \Gamma [-j+s']}\,. \label{finalana}
\end{align}
The fundamental mode, $n=0$, for the axial dipole ($j=2\,,s'=3$) reads 
\be
M\omega_I\approx (\tilde{a}-2r_+\mu)\frac{40 (M\mu) ^{11}}{19683}\,.
\ee
The formula above is valid when $0\neq\tilde{a}\ll M\mu$ whereas, in the nonrotating case, it must be multiplied by a factor $2$ as explained above. A comparison with the numerical results for the non-rotating case and for the rotating case is shown in Fig.~\ref{fig:axial_ana} and in Fig.~\ref{fig:ana_vs_num}, respectively.



We note the importance of the factor $\Gamma [j+i P-\sigma +1]$, which takes the form $\Gamma [j-s'+1]$ at lowest order in $P$ and $\tilde{a}/\epsilon$ and diverges because $s'>j$ $(s'=3,j=2)$. This is not the case in the axial perturbations of the Proca field $(s'=1,j\geq 1)$ and the perturbations of the scalar field $(s'=0,j\geq 0)$~\cite{Detweiler:1980uk,Furuhashi:2004jk,Rosa:2011my,Pani:2012bp}.  It is this factor that contributes with a term $(M\mu)^{-2S}$ for the imaginary part of the quasibound frequency, resulting in a power-law of the form $\omega_I/\mu\propto -(M\mu)^{4j-2S+5}=-(M\mu)^{4l+2S+5}$.

\subsection{Note on the monopole of Proca and massive spin-2 field}
\begin{figure}[htb]
\begin{center}
\epsfig{file=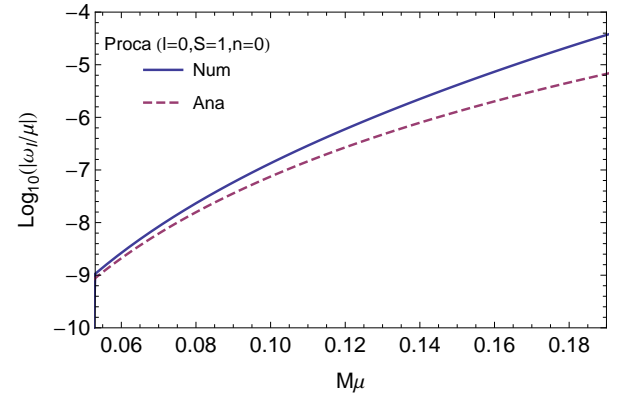,width=7.9cm,angle=0,clip=true}
\caption{Comparison between the numerical and analytical results for the Proca field mode $l=0$, $n=0$ as a function of the mass coupling $M\mu$. The solid line shows the numerical data and the dashed shows the analytical formula $\omega_I/\mu\approx -\frac{3}{4}(M\mu)^{7}$.\label{fig:proca_mono}}
\end{center}
\end{figure}

The monopole equation for the Proca field~\cite{Rosa:2011my} is given by
\be
\frac{d^2 u_{(2)}}{dr_*^2}+\left[\omega^2-f(r)\left(\mu^2+\frac{2}{r^2}-\frac{6M}{r^3}\right)\right]u_{(2)}=0\,.
\ee
This can be written in the form~\eqref{di_ana} taking $\tilde{a}=0$, $j=l+S=1$, and $s'=2$. We can then solve analytically this equation in the same way as we did for the axial dipole and all the formulas apply. 
%
%
We then find that for this mode
\be
\frac{\omega_I}{\mu}\approx -\frac{8 (M\mu) ^7(n+1) (n+3)}{(n+2)^5} \,,
\ee
in agreement with the numerical results of Rosa and Dolan~\cite{Rosa:2011my}. In Fig.~\ref{fig:proca_mono} we compare the numerical results and the analytical formula in the small-mass limit.

Another interesting behavior that we can infer comparing with the axial dipole is that it seems that $s'$ is simply given by the sum of the spin projection $S$ and the spin of the field, i.e., $s'=s+S=1+1=2$ for the monopole of the Proca field and $s'=s+S=2+1=3$ for the massive spin-2 field. 

Unfortunately the monopole equation for the massive spin-2 field~\eqref{evenl0} does not have a simple and understandable form in the limit $z\ll 1$ due to the complex form of the potential. However in the limit $z\gg 1$ we can deduce the equation 
\be
\frac{d^2}{dx^2}\left(x R_0\right)+\left[-\frac{1}{4}+\frac{\nu}{x}-\frac{6}{x^2}\right]xR_0=0\,,
\ee
where $R_0=\varphi_0/r$. This looks exactly like the axial dipole equation in the same limit~\eqref{di_zgg1}. By comparison we can see that the monopole acquires a centrifugal term with $j=l+S=2$ in agreement with our numerical results.

\bibliography{ref}  

\end{document}